\documentclass[12pt,twoside, a4paper]{article}
\def\pd{\partial}
\def\mc{\mathcal}

\usepackage[dvips]{graphicx}
\usepackage{amssymb}
\usepackage{amssymb,amsmath}
\usepackage{graphicx}
\usepackage{dsfont}
\usepackage{caption}
\usepackage{subcaption}
\input{epsf.sty}  
\begin{document}
\begin{center}
\LARGE{\textbf{$dS_4$ vacua from matter-coupled 4D $N=4$ gauged supergravity}}
\end{center}
\vspace{1 cm}
\begin{center}
\large{\textbf{H. L. Dao}$^a$ and \textbf{Parinya Karndumri}$^b$}
\end{center}
\begin{center}
$^a$\textsl{Department of Physics, National University of Singapore,
3 Science Drive 2, Singapore 117551}\\
E-mail: hl.dao@u.nus.edu\\
$^b$\textsl{String Theory and Supergravity Group, Department
of Physics, Faculty of Science, Chulalongkorn University, 254 Phayathai Road, Pathumwan, Bangkok 10330, Thailand} \\
E-mail: parinya.ka@hotmail.com \vspace{1 cm}
\end{center}
\begin{abstract}
We study $dS_4$ vacua within matter-coupled $N=4$ gauged supergravity in the embedding tensor formalism. We derive a set of conditions for the existence of $dS_4$ solutions by using a simple ansatz for solving the extremization and positivity of the scalar potential. We find two classes of gauge groups that lead to $dS_4$ vacua. One of them consists of gauge groups of the form $G_{\textrm{e}}\times G_{\textrm{m}}\times H$ with $H$ being a compact group and $G_{\textrm{e}}\times G_{\textrm{m}}$ a non-compact group with $SO(3)\times SO(3)$ subgroup and dynonically gauged. These gauge groups are the same as those giving rise to maximally supersymmetric $AdS_4$ vacua. The $dS_4$ and $AdS_4$ vacua arise from different coupling ratios between $G_{\textrm{e}}$ and $G_{\textrm{m}}$ factors. Another class of gauge groups is given by $SO(2,1)_{\textrm{e}}\times SO(2,1)_{\textrm{m}}\times G_{\textrm{nc}}\times G'_{\textrm{nc}}\times H$ with $SO(2,1)$, $G_{\textrm{nc}}$ and $G'_{\textrm{nc}}$ dyonically gauged. We explicitly check that all known $dS_4$ vacua in $N=4$ gauged supergravity satisfy the aforementioned conditions, hence the two classes of gauge groups can accommodate all the previous results on $dS_4$ vacua in a simple framework. Accordingly, the results provide a new approach for finding $dS_4$ vacua. In addition, relations between the embedding tensors for gauge groups admitting $dS_4$ and $dS_5$ vacua are studied, and a new gauge group, $SO(2,1)\times SO(4,1)$, with a $dS_4$ vacuum is found by applying these relations to $SO(1,1)\times SO(4,1)$ gauge group in five dimensions.
\end{abstract}
\newpage
\tableofcontents
\newpage
%%%%%%%%%%%%%%%%%%%%%%%%%%%%%%%%%%%%%%%%%%%%%%%%%%%%%%%%%%%%%%%%%%%%%%%%%%%%%%%%%%%%%%%%%%%%%%%%%%%%%%%%%%%%%%%%%%%%%%%%%%%%%%%%%%%%%%%%%
\section{Introduction}
De Sitter $(dS)$ vacua are solutions of general relativity and gauged supergravity with positively constant curvature. Although these solutions are originally of mathematical interest, cosmological observations, see for example \cite{dS_cosmo1,dS_cosmo2,dS_cosmo3}, suggest that the universe has a very small positive value of cosmological constant. Furthermore, these solutions have attracted much attention during the past twenty years due to the proposed dS/CFT correspondence \cite{dS_CFT}, a holographic duality between a theory of gravity on $dS$ background and a Euclidean conformal field theory along the line of the AdS/CFT correspondence \cite{maldacena}. 
\\
\indent Unlike the $AdS$ counterpart found naturally in many gauged supergravities, $dS$ vacua are very rare and (if they exist) the embedding in string/M-theory is highly non-trivial. Various approaches have been devoted to search for these vacua with only a small number of solutions found, see \cite{RS,dS1,dS2,dS3,dS4,flux_moduli,complete_Mink_vacua,dS_string6,dS6,dS_Sdual,dS8,dS9,dS_string1,dS_string2,dS_string3,dS_string3_1,dS_string4,dS_string5} for an incomplete list. All these results even suggest that string/M-theory does not admit de Sitter solutions, for a recent review see \cite{Danielsson-18} and references therein. In addition, there are a number of previous works considering de Sitter solutions of gauged supergravities. In four dimensions, de Sitter vacua are extensively studied, see for example \cite{Hull-84,de-Roo-Panda-1,Trigiante,de-Roo-Panda-2,Roest-09}. On the other hand, de Sitter vacua in higher dimensions are less known \cite{smet-PhD,gunaydin-00,ogetbil-1,ogetbil-2,smet-05,dS7,dS5}.  
\\
\indent In this paper, we study $dS_4$ vacua in four-dimensional $N=4$ gauged supergravity coupled to vector multiplets constructed in \cite{N4_gauged_SUGRA} using the embedding tensor formalism, see \cite{Eric_N4_4D,de_Roo_N4_4D} for an earlier construction.
We do not attempt to find new $dS_4$ solutions but to introduce a new approach for finding $dS_4$ vacua. We will extend a recent result initiated in the study of $dS_5$ vacua in five-dimensional $N=4$ gauged supergravity given in \cite{dS5}. Unlike the previous results on $dS_4$ solutions mostly obtained from using the old construction of \cite{de_Roo_N4_4D}, working in the embedding tensor formalism has the advantage that different deformations, gaugings and non-trivial $SL(2)$ phases, are encoded in a single framework. Furthermore, an explicit form of the gauge group under consideration needs not be specified at the beginning. This allows to formulate a general setup and subsequently apply the results to a particular gauge group.
\\
\indent We now describe the procedure used in our analysis. In general, the scalar potential of a gauged supergravity can be written as a quadratic function of fermion-shift matrices. In \cite{dS5}, the extremization and positivity of the scalar potential are solved by using a particular form of an ansatz such that the gravitino-shift matrix (usually denoted by the $A_1$ tensor) vanishes. This guarantees the positivity of the potential and, with a suitable condition, the potential can be extremized. With the help of the embedding tensor formalism, a general form of gauge groups that lead to $dS_4$ vacua can be determined from the conditions imposed on the fermion-shift matrices. 
\\
\indent It should be noted that the procedure and the resulting conditions are very similar to those arising from the existence of supersymmetric $AdS_4$ vacua given in \cite{AdS4_N4_Jan}. However, there is a crucial difference in the sense that the conditions for $dS_4$ are derived from a particular ansatz, but those for $AdS_4$ are obtained by requiring unbroken supersymmetry. While the latter guarantee that the results are vacuum solutions of the $N=4$ gauged supergravity and, in particular, extremize the scalar potential, we need to explicitly impose the extremization of the potential in the former case. We will see that some of these extra conditions are already implied by the quadratic constraint. The remaining ones imply that the gauge groups must be dyonically embedded in the global symmetry group similar to the $AdS_4$ case.  
\\
\indent The paper is organized as follows. In section \ref{N4_SUGRA},
we review relevant formulae for computing the scalar potential of $N=4$ gauged supergravity coupled to vector multiplets in the embedding tensor formalism. In section \ref{dS4}, we derive the conditions for the scalar potential to admit $dS_4$ vacua by solving the extremization and positivity of the potential. A general form of gauge groups is also determined by solving these conditions subject to the quadratic constraint. In section \ref{dS4_example}, we explicitly verify that all the previously known $dS_4$ vacua in $N=4$ gauged supergravity are encoded in our results. Some relations between four- and five-dimensional embedding tensors for gauge groups that admit de Sitter vacua are given in section \ref{dS4_dS5}. Conclusions and comments on the results are given in section \ref{conclusions}. We also include an appendix collecting useful identities for $SO(6)$ gamma matrices.  

%%%%%%%%%%%%%%%%%%%%%%%%%%%%%%%%%%%%%%%%%%%%%%%%%%%%%%%%%%%%%%%%%%%%%%%%%%%%%%%%%%%%%%%%%%%%%%%%%%%%%%%%%%%%%%%%%%%%%%%%%%%%%%%%%%%%%%%%%
\section{Four-dimensional $N=4$ gauged supergravity coupled to vector multiplets}\label{N4_SUGRA} 
In this section, we breifly review $N=4$ gauged supergravity coupled to an arbitrary number $n$ of vector multiplets. We will mainly give relevant formulae for finding the scalar potential which is the most important part in our analysis. For more detail, interested readers are referred to \cite{N4_gauged_SUGRA}. 
\\
\indent For $N=4$ supersymmetry, there are two types of supermultiplets, gravity and vector multiplets. The former contains the graviton $e^{\hat \mu}_\mu$, four gravitini $\psi_{\mu i}$, $i=1,...,4$, six vectors $A^m_\mu$, $m=1,...,6$, four spin-$\frac{1}{2}$ fermions $\lambda_i$, and a complex scalar $\tau$. The field content of the latter is given by a vector field $A_\mu$, four gaugini $\lambda^{i}$ and six scalars $\phi^{m}$. We use the following conventions on various types of indices. Spacetime and tangent space indices will be denoted by $\mu, \nu, \ldots = 0,1,2,3$ and $\hat\mu, \hat \nu,\ldots = 0,1,2,3$, respectively. Indices $m,n,\ldots=1,2,\ldots, 6$ label vector representation of $SO(6)\sim SU(4)$ R-symmetry while $i,j,\ldots$ denote chiral spinor of $SO(6)$ or fundamental representation of $SU(4)$. The $n$ vector multiplets are labeled by indices $a,b,\ldots =1,...,n$. Accordingly, the fied content of the vector multiplets can be collectively written as $(A^a_\mu,\lambda^{ai},\phi^{am})$. 
\\
\indent The $(6n+2)$ scalars from both the gravity and vector multiplets are described by the coset manifold
\begin{eqnarray}
\mathcal M = \frac{SL(2)}{SO(2)}\times \frac{SO(6,n)}{SO(6)\times SO(n)}\, .
\end{eqnarray}
The first factor is parametrized by a complex scalar $\tau$ consisting of the dilaton $\phi$ and the axion $\chi$ from the gravity multiplet. The second factor incorporates the $6n$ scalars from the vector multiplets. A useful parametrization for these two coset manifolds is given in term of the coset representatives. For $SL(2)/SO(2)$, we will use the following form of the coset representative 
\begin{eqnarray}
\mc V_\alpha = \frac{1}{\sqrt{\textrm{Im}\tau}} \begin{pmatrix}\tau \\ 1\end{pmatrix}
\end{eqnarray}
with an index $\alpha = (+,-)$ denoting the $SL(2)$ fundamental representation. $\mc{V}_\alpha$ satisfies the relation
\begin{equation}
M_{\alpha\beta}=\textrm{Re}(\mc{V}_\alpha\mc{V}^*_\beta)\qquad \textrm{and}\qquad \epsilon_{\alpha\beta}=\textrm{Im}(\mc{V}_\alpha\mc{V}^*_\beta)
\end{equation}
in which $M_{\alpha\beta}$ is a symmetric matrix with unit determinant. $\epsilon_{\alpha\beta}$ is anti-symmetric with $\epsilon_{+-}=\epsilon^{+-}=1$. 
\\
\indent For $SO(6,n)/SO(6)\times SO(n)$, we use the coset representative ${\mc {V}_M}^A$ transforming under the global $SO(6,n)$ and local $SO(6)\times SO(n)$ symmetry by left and right multiplications, respectively. The local index $A$ can be split as $A=(m,a)$ with $m=1,2,\ldots,6$ and $a=1,2,\ldots ,n$ denoting vector representations of $SO(6)\times SO(n)$. The coset representative can then be written as
\begin{eqnarray}
{\mc V_M}^A = ({\mc V_M}^m, {\mc V_M}^a)\, .
\end{eqnarray}
Since ${\mc V_M}^A$ is an $SO(6,n)$ matrix, it satisfies the following relation
\begin{eqnarray}
\eta_{MN} = -{\mc V_M}^m {\mc V_N}^m + {\mc V_M}^a {\mc V_N}^a
\end{eqnarray}
where $\eta_{MN} = \textrm{diag}\left(-1,-1,-1,-1,-1,-1, 1,\ldots,1\right)$ is the $SO(6,n)$ invariant tensor. $\eta_{MN}$ and its inverse $\eta^{MN}$ can be used to lower and raise $SO(6,n)$ indices, $M,N,\ldots $. The $SO(6,n)/SO(6)\times SO(n)$ coset can also be described by a symmetric matrix 
\begin{equation}
M_{MN} ={\mc V_M}^m {\mc V_N}^m + {\mc V_M}^a {\mc V_N}^a
\end{equation}
which is manifestly $SO(6)\times SO(n)$ invariant.
\\
\indent Gaugings of the matter-coupled $N=4$ supergravity can be efficiently implemented by using the embedding tensor formalism \cite{N4_gauged_SUGRA}. This constant tensor describes the embedding of a gauge group $G_0$ in the global symmetry group $SL(2)\times SO(6,n)$. It turns out that $N=4$ supersymmetry allows only the following components of the embedding tensor $\xi_{\alpha M}$ and  $f_{\alpha MNP}=f_{\alpha[MNP]}$. To describe a closed subalgebra of $SL(2)\times SO(6,n)$, the embedding tensor needs to satisfy the quadratic constraints
\begin{eqnarray}
{\xi_\alpha}^M\xi_{\beta M}=0,\qquad \epsilon^{\alpha\beta}({\xi_\alpha}^Pf_{\beta_{MNP}}+\xi_{\alpha M}\xi_{\beta N})&=&0,\nonumber \\
{\xi_{(\alpha}}^Pf_{\beta) MNP}=0,\qquad 3f_{\alpha R[MN}{f_{\beta PQ]}}^R+2\xi_{(\alpha [M}f_{\beta)NPQ]}&=&0,\nonumber \\
\epsilon^{\alpha\beta}\left(f_{\alpha MNR}{f_{\beta PQ}}^R-{\xi_\alpha}^Rf_{\beta R[M[P}\eta_{Q]N}-\xi_{\alpha [M}f_{\beta N]PQ}+\xi_{\alpha [P}f_{\beta Q]MN}\right)&=&0\, .\label{QC}
\end{eqnarray}
\indent In this paper, we are only interested in maximally symmetric solutions of $N=4$ gauged supergravity with only the metric and scalars non-vanishing. Therefore, we will set all the other fields to zero from now on. In this case, the bosonic Lagrangian takes the form of
\begin{eqnarray}
e^{-1} \mc L = \frac{1}{2}R + \frac{1}{16} \pd_\mu M_{MN} \pd^\mu M^{MN} - \frac{1}{4(\text{Im}\tau)^2} \pd_\mu \tau \pd^\mu \tau^* - V\, .
\end{eqnarray}
The scalar potential $V$ reads 
\begin{eqnarray}
V&=&\frac{g^2}{16}\left[f_{\alpha MNP}f_{\beta
QRS}M^{\alpha\beta}\left[\frac{1}{3}M^{MQ}M^{NR}M^{PS}+\left(\frac{2}{3}\eta^{MQ}
-M^{MQ}\right)\eta^{NR}\eta^{PS}\right]\right.\nonumber \\
& &\left.-\frac{4}{9}f_{\alpha MNP}f_{\beta
QRS}\epsilon^{\alpha\beta}M^{MNPQRS}+3{\xi_\alpha}^M{\xi_\beta}^NM^{\alpha\beta}M_{MN}\right]
\end{eqnarray}
where $M^{MN}$ and $M^{\alpha\beta}$ are the inverse matrices of $M_{MN}$ and $M_{\alpha\beta}$, respectively. $M^{MNPQRS}$ is obtained by raising indices of $M_{MNPQRS}$ defined by
\begin{equation}
M_{MNPQRS}=\epsilon_{mnpqrs}\mc{V}_{M}^{\phantom{M}m}\mc{V}_{N}^{\phantom{M}n}
\mc{V}_{P}^{\phantom{M}p}\mc{V}_{Q}^{\phantom{M}q}\mc{V}_{R}^{\phantom{M}r}\mc{V}_{S}^{\phantom{M}s}\, .\label{M_6}
\end{equation}
\indent In subsequent analysis, it is useful to rewrite the potential in terms of the fermion-shift matrices $A_1^{ij}$, $A_2^{ij}$ and ${A_{2ai}}^j$ that appear in the fermion mass-like terms and supersymmetry transformations of fermions. In general, the scalar potential can be determined by the supersymmetric Ward identity
\begin{equation}
\frac{1}{4}\delta^i_jV=\frac{1}{2}{A_{2aj}}^k{A^*_{2ak}}^i+\frac{1}{9}A^{ik}_2A^*_{2jk}-\frac{1}{3}A^{ik}_1A^*_{1jk}
\end{equation}
which, after a contraction of indices, gives
\begin{eqnarray}
V = \frac{1}{2}{A_{2ai}}^j {A^*_{2aj}}^i + \frac{1}{9}A_2^{ij}A^*_{2ij} -\frac{1}{3}A^{ij}_1 A^*_{1ij}\, .
\end{eqnarray}
In terms of the scalar coset representatives, the fermion shift-matrices are given by
\begin{eqnarray}
A^{ij}_1 &=& \epsilon^{\alpha\beta}(\mc V_\alpha)^* {\mc {V}_{kl}}^M {\mc {V}_N}^{ik}{\mc V_P}^{jl} {f_{\beta M}}^{NP}, \nonumber\\
A^{ij}_2 &=& \epsilon^{\alpha\beta} \mc V_\alpha {\mc {V}_{kl}}^M {\mc V_N}^{ik} {\mc V_P}^{jl} {f_{\beta M}}^{NP}+\frac{3}{2}\epsilon^{\alpha\beta}\mc{V}_\alpha{\mc{V}_M}^{ij}{\xi_\beta}^M, \nonumber\\
{A_{2ai}}^j &=& \epsilon^{\alpha\beta} \mc V_\alpha {\mc V_a}^M {\mc V_{ik}}^N {\mc V_P}^{jk} {f_{\beta MN}}^{P}-\frac{1}{4}\delta^j_i\epsilon^{\alpha\beta}\mc{V}_\alpha {\mc{V}_a}^M\xi_{\beta M}\, .\label{definition}
\end{eqnarray}
${\mc{V}_M}^{ij}$ is obtained from ${\mc{V}_M}^m$ by converting the $SO(6)$ vector index to an anti-symmetric pair of $SU(4)$ fundamental indices using the chiral $SO(6)$ gamma matrices.

%%%%%%%%%%%%%%%%%%%%%%%%%%%%%%%%%%%%%%%%%%%%%%%%%%%%%%%%%%%%%%%%%%%%%%%%%%
\section{de Sitter vacua of $N=4$ four-dimensional gauged supergravity}\label{dS4}
In this section, we will look for gauge groups that lead to de Sitter vacua. The analysis is similar to that given in \cite{dS5} for $dS_5$ vacua. Furthermore, the procedure is closely parallel to the case of maximally supersymmetric $AdS_4$ vacua given in \cite{AdS4_N4_Jan}. 
\\
\indent Denoting the chiral $SO(6)$ gamma matrices by $\Gamma^{ij}_m$, we can write ${\mc{V}_M}^{ij}$ in term of the coset representative ${\mc{V}_M}^m$ as
\begin{equation}
{\mc{V}_M}^{ij}={\mc{V}_M}^{m}\Gamma^{ij}_m\, .
\end{equation}
Similarly, the inverse element ${\mc{V}_{ij}}^M$ is given by
\begin{equation}
{\mc{V}_{ij}}^M={\mc{V}_m}^M(\Gamma^{ij}_m)^*\, .
\end{equation}
\indent In order to have $dS_4$ vacua, we require that
\begin{equation}
\langle \delta V\rangle=0\qquad \textrm{and}\qquad \langle V\rangle>0\label{general_dS5_con}
\end{equation}
where, as in \cite{AdS4_N4_Jan}, the bracket $\langle \,\,\, \rangle$ indicates that the quantity inside is evaluated at the vacuum. In terms of the fermion-shift matrices, these conditions read
\begin{eqnarray}
\langle \delta V \rangle&=&-\frac{1}{3}\langle \delta A^{ij}_1A^*_{1ij}\rangle -\frac{1}{3}\langle \delta A^*_{1ij}A^{1ij}\rangle +\frac{1}{9}\langle \delta A_2^{ij} A^*_{2ij}\rangle \nonumber \\
& &\qquad +\frac{1}{9}\langle \delta A^*_{2ij} A_2^{ij}\rangle+\frac{1}{2}\langle\delta {A_{2ai}}^j{A^*_{2aj}}^i \rangle +\frac{1}{2}\langle {A_{2ai}}^j\delta {A^*_{2aj}}^i \rangle=0,\quad \label{extremization}\\
\textrm{and}\quad & &\frac{1}{9}\langle A^{ij}_2A^*_{2ij}\rangle+\frac{1}{2}\langle {A_{2ai}}^j{A^*_{2aj}}^i\rangle>\frac{1}{3}\langle A^{ij}_1A^*_{1ij}\rangle \, .\label{positivity}
\end{eqnarray}
In general, there are various solutions to these conditions. However, as in five dimensions, we will consider only the following two possibilities:
\begin{enumerate}
\item $\langle A_1^{ij} \rangle=\langle {A_{2ai}}^j \rangle=0$ and $\langle A^{ik}_2 A^*_{2jk} \rangle=\frac{9}{4}|\mu|^2\delta^i_j$ with $A^*_{2ij}\delta A^{ij}_2+\delta A^*_{2ij} A^{ij}_2=0$.
\item $\langle A_1^{ij} \rangle=\langle A^{ij}_2 \rangle=0$ and $\langle {A_{2ai}}^k {A^*_{2ak}}^j \rangle=\frac{1}{2}|\mu|^2\delta^j_i$ with $\delta{A_{2ai}}^j {A^*_{2aj}}^i+{A_{2ai}}^j\delta {A^*_{2aj}}^i=0$.
\end{enumerate} 
$|\mu|^2=V_0$ denotes the cosmological constant. We now consider these two sets of conditions separately. 

%%%%%%%%%%%%%%%%%%%%%%%%%%%%%%%%%%%%%%%%%%%%%%%%%%%%%%%%%%%%%%%%%%%%%%%
\subsection{$\langle A_1^{ij} \rangle=\langle {A_{2ai}}^j \rangle=0$ and $\langle A^{ik}_2 A^*_{2jk} \rangle=\frac{9}{4}|\mu|^2\delta^i_j$}\label{dS4_1}
We begin with the $\langle {A_{2ai}}^j \rangle=0$ condition. From equation \eqref{definition}, we see that the first term in $\langle {A_{2ai}}^j \rangle=0$ is traceless in $i$ and $j$ indices while the second term is the trace part. They must then vanish separately and lead to the following condition
\begin{equation}
\xi_{\alpha a}=0\label{xi_alpha_M1}
\end{equation}
where ${\xi_\alpha}^a=\langle {\mc{V}_M}^a \rangle {\xi_\alpha}^M$. Using the first condition in the quadratic constraint \eqref{QC} and ${\xi_\alpha}^m=\langle {\mc{V}_M}^m \rangle {\xi_\alpha}^M$, we find
\begin{equation}
{\xi_\alpha}^M\xi_{\beta M}=-{\xi_\alpha}^m{\xi_\beta}^m+{\xi_\alpha}^a{\xi_\beta}^a=-{\xi_\alpha}^m{\xi_\beta}^m=0
\end{equation}
for any values of $\alpha$ and $\beta$. This implies that $\xi_{\alpha m}=0$. Together with \eqref{xi_alpha_M1}, we then have
\begin{equation}
\xi_{\alpha M}=0\, .
\end{equation}
This condition implies that the gauge group is entirely embedded in $SO(6,n)$.
\\
\indent With $\xi_{\alpha M}=0$, the $\langle {A_{2ai}}^j \rangle=0$ condition gives  
\begin{equation}
\epsilon^{\alpha\beta} \langle \mc V_\alpha {\mc V_a}^M {\mc V_{ik}}^N {\mc V_P}^{jk}\rangle {f_{\beta MN}}^{P}=0\, .\label{A2a_con}
\end{equation}
In subsequent anlysis, as in \cite{AdS4_N4_Jan}, it is useful to introduce ``dressed'' complex components of the embedding tensor
\begin{equation}
\mathfrak{f}_{ABC}=\mathfrak{f}_{1\, ABC}+i\mathfrak{f}_{2\,ABC}=\langle {\mc{V}_A}^M{\mc{V}_B}^N{\mc{V}_C}^P\mc{V}_\alpha\rangle \epsilon^{\alpha\beta}f_{\beta NMP}\, .
\end{equation}
The real and imaginary parts are given explicitly by
\begin{eqnarray}
\mathfrak{f}_{1\, ABC}&=&\frac{1}{\sqrt{\langle \textrm{Im}\tau\rangle}}\langle {\mc{V}_A}^M{\mc{V}_B}^N{\mc{V}_C}^P\rangle[\langle \textrm{Re} \tau\rangle f_{-NMP}-f_{+NMP}],\\
\mathfrak{f}_{2\, ABC}&=&\langle \sqrt{\textrm{Im}\tau}{\mc{V}_A}^M{\mc{V}_B}^N{\mc{V}_C}^P\rangle f_{-NMP}\, .
\end{eqnarray}
It should also be noted that by working at the origin of the scalar manifold $SL(2)/SO(2)$ with $\textrm{Im}\tau=1$ and $\textrm{Re}\tau=0$, we see that $\mathfrak{f}_{1\, ABC}$ and $\mathfrak{f}_{2\, ABC}$ correspond to electric and magnetic components of $f_{\alpha MNP}$, respectively. In particular, we have for $\xi_{\alpha M}=0$,
\begin{eqnarray}
A^{ij}_1&=&-\mathfrak f^*_{mnp}(\Gamma^m\Gamma^{*n}\Gamma^p)^{ij},\label{A1_f}\\
A^{ij}_2&=&-\mathfrak f_{mnp}(\Gamma^m\Gamma^{*n}\Gamma^p)^{ij},\label{A2_f}\\
{A_{2ai}}^j&=&\mathfrak f_{amn}{(\Gamma^{mn})_i}^j\, .\label{A2a_f}
\end{eqnarray}
\indent With all these ingredients, we can rewrite equation \eqref{A2a_con} as
\begin{equation}
\mathfrak{f}_{amn}{(\Gamma^{mn})_i}^{j}=0
\end{equation}
which gives
\begin{equation}
\mathfrak{f}_{amn}=0\, .
\end{equation}
This also implies that both real and imaginary parts of $\mathfrak{f}_{amn}$ vanish or equivalently
\begin{equation}
f_{\alpha amn}=0\, .
\end{equation}
The conditions $\langle A^{ij}_1\rangle=0$ and $\langle A_2^{ik}A^*_{2jk}\rangle =\frac{9}{4}|\mu|^2\delta^i_j$ give
\begin{equation}
\mathfrak{f}_{mnp}(\Gamma^n\Gamma^{*m}\Gamma^p)^{ij}=-\sqrt{3}\mu P^{ij}\qquad \textrm{and}\qquad \mathfrak{f}^*_{mnp}(\Gamma^n\Gamma^{*m}\Gamma^p)^{ij}=0 
\end{equation}
where we have introduced a constant matrix $P^{ij}$ with the property $P^{ik}P^*_{jk}=\delta^i_j$. 
\\
\indent It can be seen that all of these conditions are very similar to those for the existence of supersymmetric $AdS_4$ vacua with $\mathfrak{f}_{mnp}$ and $\mathfrak{f}_{mnp}^*$ interchanged and $\mu$ replaced by $\sqrt{3}\mu$. Using the identity given in \eqref{3_Gamma}, we obtain  
\begin{equation}
\mathfrak{f}^*_{mnp}\mathfrak{f}^{mnp}=6|\mu|^2\qquad \textrm{and}\qquad \mathfrak{f}^*_{mnp}+i\epsilon_{mnpqrs}\mathfrak{f}^*_{qrs}=0\, . \label{f_fc}
\end{equation}
In terms of real and imaginary parts, these can be written as
\begin{eqnarray}
& &\mathfrak{f}_{1\,mnp}{\mathfrak{f}_1\, }^{mnp}+\mathfrak{f}_{2\,mnp}{\mathfrak{f}_2\, }^{mnp}=6|\mu|^2,\\
& &\mathfrak{f}_{1\,mnp}=\epsilon_{mnpqrs}\mathfrak{f}_{2\,qrs}\qquad \textrm{and}\qquad \mathfrak{f}_{2\,mnp}=-\epsilon_{mnpqrs}\mathfrak{f}_{1\,qrs}\label{f1_f2_con}
\end{eqnarray}
which imply that both $\mathfrak f_1$ and $\mathfrak f_2$ cannot be zero, hence the gauge groups are essentially dyonic. 
\\
\indent All these conditions must be solved subject to the quadratic constraint which for $\xi_{\alpha M}=0$ simplifies considerably
\begin{eqnarray}
f_{\alpha R[MN}{f_{\beta PQ]}}^R = 0 \qquad \textrm{and}\qquad \epsilon^{\alpha\beta}f_{\alpha MNR} {f_{\beta PQ}}^R = 0\, .
\end{eqnarray} 
In terms of $\mathfrak{f}_{ABC}$, these constraints read
\begin{equation}
{\mathfrak{f}_{[AB}}^E\mathfrak{f}_{CD]E}=0,\qquad \textrm{Re}({\mathfrak{f}_{[AB}}^E\mathfrak{f}^*_{CD]E})=0,\qquad \textrm{Im}({\mathfrak{f}_{AB}}^E\mathfrak{f}^*_{CDE})=0\, .\label{complex_QC}
\end{equation}
\indent It has been shown in \cite{AdS4_N4_Jan} that the conditions \eqref{f_fc} and the $(mnpq)$-components of the quadratic constraint \eqref{complex_QC} has a unique solution of the form
\begin{equation}
\mathfrak{f}_{123}=\frac{1}{\sqrt{2}}\mu\qquad \textrm{and}\qquad \mathfrak{f}_{456}=\frac{i}{\sqrt{2}}\mu \label{sol_1}
\end{equation}  
or, equivalently,
\begin{equation}
\mathfrak{f}_{1\,123}=\frac{1}{\sqrt{2}}\mu\qquad \textrm{and}\qquad \mathfrak{f}_{2\, 456}=\frac{1}{\sqrt{2}}\mu\, .\label{sol_2}
\end{equation}  
Note some numerical changes especially the different relative sign between $\mathfrak{f}_{1\,123}$ and $\mathfrak{f}_{2\,456}$ which is opposite to that of the $AdS_4$ case.  
\\
\indent At this point, all the remaining parts of the whole analysis are essentially the same as in \cite{AdS4_N4_Jan}. In particular, the resulting gauge groups that can give rise to $dS_4$ vacua must take the same form as in the $AdS_4$ case. We will not repeat all the details here but simply summarize the structure of possible gauge groups. First of all, it should be noted that other components of the embedding tensors, $\mathfrak{f}_{mab}$ and $\mathfrak{f}_{abc}$, are not constrained by the existence of $dS_4$ vacua.
\\
\indent For $\mathfrak{f}_{mab}=\mathfrak{f}_{abc}=0$, the gauge group is only generated by $\mathfrak{f}_{mnp}$. With the solution \eqref{sol_1}, we find the gauge group of the form
\begin{equation}
SO(3)_{\textrm{e}}\times SO(3)_{\textrm{m}}
\end{equation}
which can be embedded entirely in the $SO(6)$ R-symmetry. The full gauge group is dyonically gauged by the six graviphotons with the two $SO(3)$ factors being electrically and magnetically gauged, respectively.
\\
\indent For $\mathfrak{f}_{mab}=0$ but $\mathfrak{f}_{abc}\neq 0$, $\mathfrak{f}_{mmp}$ and $\mathfrak{f}_{abc}$ lead to gauge groups of the form
\begin{equation}
SO(3)_{\textrm{e}}\times SO(3)_{\textrm{m}}\times G_0^v\subset SO(6,n)\, .
\end{equation}
with $G_0^v$ being a compact group gauged by vector fields from the vector multiplets.  
\\
\indent For $\mathfrak{f}_{mab}\neq 0$ and $\mathfrak{f}_{abc}\neq 0$, there can be two subsets of $\mathfrak{f}_{abc}$ in which one has common indices with $\mathfrak{f}_{mab}$, but the other one does not. The latter again forms a separate compact group. The $(mnpq)$-components of the quadratic constraint implies that $\mathfrak{f}_{1\,mab}$ and $\mathfrak{f}_{2\, mab}$ are non-vanishing only for $m=1,2,3$ and $m=4,5,6$, respectively. Therefore, $\mathfrak{f}_{mab}$ extend $SO(3)_{\textrm{e}}\times SO(3)_{\textrm{m}}$ to a product of non-compact groups G$_{\textrm{e}}\times G_{\textrm{m}}$ containing $SO(3)_{\textrm{e}}\times SO(3)_{\textrm{m}}$ as a subgroup. The general form of gauge groups is then given by 
\begin{equation}
G_{\textrm{e}}\times G_{\textrm{m}}\times G_0^v\subset SO(6,n)\, .
\end{equation}
\indent Finally, we will explicitly check that solutions to all of the above conditions indeed extremize the scalar potential. This is crucial because our conditions do not arise from the requirement of supersymmetric vacua as in the $AdS_4$ case. To proceed, we first note a number of useful relations
\begin{eqnarray}
\delta{\mc{V}_M}^m&=&{\mc{V}_M}^a\delta \phi^{ma},\qquad \delta{\mc{V}_M}^a={\mc{V}_M}^m\delta \phi^{ma}, \nonumber \\
\delta \mc{V}_\alpha &=&\frac{i}{2\textrm{Im}\tau}\left(\mc{V}_\alpha^*\delta\tau -\mc{V}_\alpha \delta \textrm{Re}\tau\right) 
\end{eqnarray}
from which it follows that
\begin{eqnarray}
\delta \mathfrak{f}_{npq}&=&-3\delta_{m[n}\mathfrak{f}_{pq]a}\delta \phi^{ma}+\frac{1}{\textrm{Im}\tau}\textrm{Im}\mathfrak{f}_{npq}\delta \textrm{Re}\tau-\frac{1}{2}\textrm{Im}\tau \mathfrak{f}^*_{npq}\delta \textrm{Im}\tau,\label{delta_fmnp}\\
\delta \mathfrak{f}_{npb}&=&(2\delta_{m[n}\mathfrak{f}_{p]ab}-\delta_{ab}\mathfrak{f}_{mnp})\delta \phi^{ma}+\frac{1}{\textrm{Im}\tau}\textrm{Im}\mathfrak{f}_{npb}\delta \textrm{Re}\tau-\frac{1}{2}\textrm{Im}\tau \mathfrak{f}^*_{npb}\delta \textrm{Im}\tau\, .\label{delta_famn}
\end{eqnarray}
\indent Using \eqref{A2_f}, it is straightforward to compute
\begin{eqnarray}
\delta V&=&\frac{1}{9}\left(A_2^{ij}\delta A^*_{2ij}+\delta A^{ij}_2 A^*_{2ij}\right)\nonumber \\
&=&\frac{1}{9}\mathfrak{f}_{mnp}\delta \mathfrak{f}_{qrs}(\Gamma^m\Gamma^{*n}\Gamma^p)^{ij}(\Gamma^q\Gamma^{*r}\Gamma^s)^*_{ij}+\textrm{c.c.}
\end{eqnarray}
Using the identity \eqref{6_Gamma_trace}, we can reduce this to 
\begin{equation}
\delta V=-\frac{8}{3}\mathfrak{f}_{mnp}\delta \mathfrak{f}^{*mnp}+\textrm{c.c.}\, .
\end{equation}
Since $\mathfrak{f}_{amn}=0$ (from $\langle {A_{2ai}}^j\rangle=0$), we immediately see from \eqref{delta_fmnp} that $\langle \delta_{\phi}V\rangle =0$. 
\\
\indent Furthermore using \eqref{A1_f} and \eqref{A2_f}, we can derive the following results
\begin{equation}
\delta_{\textrm{Im}\tau}A^{ij}_1=-\frac{1}{2\textrm{Im}\tau}A^{ij}_2\delta \textrm{Im}\tau\quad\textrm{and}\quad \delta_{\textrm{Im}\tau}A^{ij}_2=-\frac{1}{2\textrm{Im}\tau}A^{ij}_1\delta \textrm{Im}\tau\, .  
\end{equation}
Using $\langle A^{ij}_1\rangle=0$, we obtain $\langle\delta_{\textrm{Im}\tau}A_2^{ij}\rangle=0$, hence $\langle\delta_{\textrm{Im}\tau}V\rangle=0$. Finally, we consider the variation with respect to $\textrm{Re}\tau$ which reads
\begin{eqnarray}
\delta_{\textrm{Re}\tau}V&=&-\frac{2}{3\textrm{Im}\tau}\textrm{Im}\mathfrak{f}_{mnp}(\mathfrak{f}^{*mnp}+\mathfrak{f}^{mnp})\delta \textrm{Re}\tau\nonumber \\
&=&-\frac{4}{3\textrm{Im}\tau}{\mathfrak{f}_1}^{mnp}\mathfrak{f}_{2\, mnp}\delta \textrm{Re}\tau\, .
\end{eqnarray}  
This vanishes due to the condition \eqref{f1_f2_con} which implies that $\mathfrak{f}_{1\, mnp}$ and $\mathfrak{f}_{2\, mnp}$ have no common indices. We then conclude that the conditions $\langle A^{ij}_1\rangle=\langle {A_{2ai}}^j\rangle =0$ and $\langle A^{ik}_2A^*_{2jk}\rangle=\frac{9}{4}|\mu|^2\delta^i_j$ extremize the scalar potential and give $dS_4$ vacua of the matter-coupled $N=4$ gauged supergravity. 
\\
\indent As a final comment, we note that although the gauge groups giving rise to $dS_4$ vacua are exactly the same as those leading to supersymmetric $AdS_4$ solutions, the two types of vacua occur at different values of the ratio between the coupling constants of $SO(3)_{\textrm{e}}$ ($G_{\textrm{e}}$) and $SO(3)_{\textrm{m}}$ ($G_{\textrm{m}}$). More precisely, the two cases have the coupling ratios with opposite sign, recall the sign change in \eqref{sol_2} as compared to the results of \cite{AdS4_N4_Jan}. We will see this in explicit examples in the next section.  

%%%%%%%%%%%%%%%%%%%%%%%%%%%%%%%%%%%%%%%%%%%%%%%%%%%%%%%%%%%%%%%%%%%%%%%
\subsection{$\langle A_1^{ij} \rangle=\langle A^{ij}_2 \rangle=0$ and $\langle {A_{2ai}}^k {A^*_{2ak}}^j \rangle=\frac{1}{2}|\mu|^2\delta^j_i$}\label{dS4_2}
We now look at another possibility for $dS_4$ vacua to exist. In this case, we require that 
\begin{equation}
\langle A_1^{ij} \rangle=\langle A^{ij}_2 \rangle=0\qquad \textrm{and}\qquad \langle {A_{2ai}}^k {A^*_{2ak}}^j \rangle=\frac{1}{2}|\mu|^2\delta^j_i\, .\label{dS4_con2}
\end{equation}
Since $A^{ij}_2$ consists of two parts, one symmetric and the other anti-symmetric in $i$ and $j$ indices, these two parts must vanish separately. Setting the anti-symmetric part to zero gives
\begin{equation} 
\xi_{\alpha m}=0\, .
\end{equation}
We again use the quadratic constraint ${\xi_\alpha}^M\xi_{\beta M}=0$ and find that
\begin{equation}
{\xi_\alpha}^a{\xi_\beta}^a=0
\end{equation}
which gives $\xi_{\alpha a}=0$. Therefore, we have $\xi_{\alpha M}=0$ as in the previous case.
\\
\indent With this result, the first two conditions in \eqref{dS4_con2} give 
\begin{equation}
\mathfrak{f}_{mnp}(\Gamma^n\Gamma^{*m}\Gamma^p)^{ij}=0\qquad \textrm{and}\qquad \mathfrak{f}^*_{mnp}(\Gamma^n\Gamma^{*m}\Gamma^p)^{ij}=0 
\end{equation}
which imply that
\begin{equation}
\mathfrak{f}=\mathfrak{f}^*=0\qquad \textrm{or}\qquad f_{\alpha mnp}=0,\label{fmnp_con2}
\end{equation}
so, in this case, compact non-abelian subgroups of the $SO(6)$ R-symmetry are not gauged. 
\\
\indent For the last condition in \eqref{dS4_con2}, a straightforward computation gives
\begin{equation}
\frac{1}{2}|\mu|^2\delta^i_j=\langle {A_{2aj}}^k{A^*_{2ak}}^i\rangle=-\frac{1}{2}\mathfrak{f}_{amn}\mathfrak{f}^*_{apq}{\{\Gamma^{mn},\Gamma^{pq}\}^i}_j\, .
\end{equation}
Using the identity \eqref{Gamma_mn_anti_com}, we arrive at
\begin{equation}
|\mu|^2\delta^i_j=-2\mathfrak{f}^{amn}\mathfrak{f}^{*apq}{(\Gamma_{mnpq})^i}_j+4\mathfrak{f}_{amn}\mathfrak{f}^{*amn}\delta^i_j
\end{equation}
which gives
\begin{equation}
\mathfrak{f}_{amn}\mathfrak{f}^{*amn}=\frac{1}{4}|\mu|^2\qquad \textrm{and}\qquad \mathfrak{f}^{a[mn}\mathfrak{f}^{*pq]a}=0\, .\label{ff_star_con}
\end{equation}
From the first condition, we see that $\mathfrak{f}_{amn}$ must be non-vanishing for the $dS_4$ vacua to exist ($\mu\neq 0$). Therefore, the gauge groups must be necessarily non-compact.
\\
\indent We then consider the extremization of the scalar potential
\begin{eqnarray}
\delta V&=&\frac{1}{2}\left(\delta {A_{2ai}}^j{A^*_{2aj}}^i+{A_{2ai}}^j\delta {A^*_{2aj}}^i\right)\nonumber \\
&=&-\frac{1}{2}\mathfrak{f}^*_{apq}\delta \mathfrak f_{amn}\textrm{Tr}\{\Gamma^{mn},\Gamma^{pq}\}+\textrm{c.c.} \, .
\end{eqnarray}
Using $\delta \mathfrak{f}_{amn}$ from \eqref{delta_famn} and the result in \eqref{fmnp_con2}, we obtain, upon setting $\delta V=0$, 
\begin{eqnarray}
\delta_\phi V=0&:&\qquad \mathfrak{f}^{*amn}\mathfrak{f}_{abm}=0,\label{ext1}\\
\delta_{\textrm{Re}\tau}V=0 &:&\qquad \textrm{Im}\mathfrak{f}_{amn}\textrm{Re}\mathfrak{f}^{amn}=0\nonumber \\ 
\textrm{or}& &\qquad \mathfrak{f}_{1\, amn}{\mathfrak{f}_2}^{amn}=0,\label{ext2}\\
\delta_{\textrm{Im}\tau}V=0&:&\qquad \mathfrak{f}_{amn}\mathfrak{f}^{amn}=0\nonumber \\
\textrm{or}& & \qquad \mathfrak{f}_{1\, amn}{\mathfrak{f}_1}^{amn}=\mathfrak{f}_{2\, amn}{\mathfrak{f}_2}^{amn}\qquad \textrm{and} \qquad \mathfrak{f}_{1\, amn}{\mathfrak{f}_2}^{amn}=0\, . \label{ext4}
\end{eqnarray}
Note that the second condition in \eqref{ext4} is the same as \eqref{ext2} and implies that $\mathfrak{f}_{1\, amn}$ and $\mathfrak{f}_{2\, amn}$ have no common indices. 
\\
\indent To determine the form of possible gauge groups, we need to solve all the above conditions subject to the quadratic constraint \eqref{complex_QC}. By substituting the results from \eqref{ext2} and \eqref{ext4} in the first condition of \eqref{ff_star_con}, we find
\begin{equation}
\mathfrak{f}_{1\, amn}{\mathfrak{f}_1}^{amn}=\mathfrak{f}_{2\, amn}{\mathfrak{f}_2}^{amn}=\frac{1}{8}|\mu|^2\, .
\end{equation}
This result and the condition \eqref{ext2} imply that, apart from being non-compact, the gauge group must be a product of at least two non-compact groups and dyonically embedded in $SO(6,n)$ since both $\mathfrak{f}_{1\, amn}$ and $\mathfrak{f}_{2\, amn}$ or $f_{\pm amn}$ are non-vanishing. 
\\
\indent Finally, using the result from \eqref{fmnp_con2}, we find that the second condition in \eqref{ff_star_con} is already implied by the $(mnpq)$-component of the quadratic constraint. Therefore, we have a set of consistent conditions to be imposed on the embedding tensor. In the following, we will look for explicit solutions and possible forms of the corresponding gauge groups. The analysis will be closely parallel to that in the previous case.  
\\
\indent We first note that components $\mathfrak{f}_{mab}$ and $\mathfrak{f}_{abc}$ are not constrained by the existence of $dS_4$ vacua. These components can be anything without affecting the $dS_4$ vacua. However, the structure of gauge groups will be different for different values of $\mathfrak{f}_{mab}$ and $\mathfrak{f}_{abc}$. We now look at various possibilities.
\\
\indent For the simplest case of $\mathfrak{f}_{mab}=\mathfrak{f}_{abc}=0$, the only non-vanishing components of the embedding tensor are given by $\mathfrak{f}_{amn}$. Since $\mathfrak{f}_{abc}=0$, the compact part of the gauge group must be an abelian $SO(2)$ group. $\mathfrak{f}_{amn}$ then leads to $SO(2,1)$ gauge group. The full gauge group generated by both real and imaginary parts of $\mathfrak{f}_{amn}$, or $f_{\pm amn}$, is given by a product of two $SO(2,1)$ factors, electrically and magnetically gauged,  
\begin{equation}
SO(2,1)_{\textrm{e}}\times SO(2,1)_{\textrm{m}}\, .
\end{equation}
In this gauge group, the two compact generators are embedded in the matter directions, $a,b=1,2,\ldots, n$. In notation of \cite{de-Roo-Panda-2}, this gauge group can be written as
\begin{equation}
SO(2,1)_+\times SO(2,1)_+\, .
\end{equation} 
It should be noted that the plus sign indicates that the $SO(2)\times SO(2)$ compact subgroup is embedded in the positive part of the $SO(6,n)$ invariant tensor $\eta_{MN}$.   
\\
\indent We then move to the case of $\mathfrak{f}_{mab}=0$ but $\mathfrak{f}_{abc}\neq 0$. The $(mnpa)$- and $(mabc)$-components of the quadratic constraint are trivially satisfied while the $(abcd)$-component reduces to the standard Jacobi's identity for $\mathfrak{f}_{abc}$ corresponding to a compact group $H_{\textrm{c}}$. The $(mnab)$-component of the quadratic constraint implies that $\mathfrak{f}_{amn}$ together with $\mathfrak{f}_{abc}$ generate a non-compact group $G_{\textrm{nc}}$. The full gauge group with both electric and magnetic factors taken into account is then given by
\begin{equation} 
G_{\textrm{nc,e(m)}}\times G'_{\textrm{nc,m(e)}}\, .
\end{equation}
It is useful to note that since $\mathfrak{f}_{1\, amn}$ and $\mathfrak{f}_{2\, amn}$ cannot have common indices, the number of non-compact generators $n_{\textrm{nc}}$ for $G_{\textrm{nc}}$ and $G'_{\textrm{nc}}$ must satisfy $2\leq n_{\textrm{nc}}\leq 4$. An example for the gauge groups with $n=6$ vector multiplets is given by
\begin{equation}
SO(2,1)_{\textrm{e(m)}}\times SU(2,1)_{\textrm{m(e)}}\qquad \textrm{or}\qquad SO(2,1)_+\times SU(2,1)_+\, .\label{SO2_1_SU2_1_group}
\end{equation}
It should be noted that, for $n=6$, there are in total $12$ vector fields, so the full gauge group also contains an additional abelian $SO(2)$ factor corresponding to the gauge symmetry of the remaining gauge field. However, matter fields are not charged under this $SO(2)$ factor as in the ungauged $N=4$ supergravity. We have accordingly omitted this factor in equation \eqref{SO2_1_SU2_1_group}.
\\
\indent We now consider the case $\mathfrak{f}_{abc}=0$ but $\mathfrak{f}_{mab}\neq 0$. For $\mathfrak{f}_{abc}=0$, we again find that $\mathfrak{f}_{amn}$ lead to $SO(2,1)$ gauge group. Furthermore, since the existence of $dS_4$ vacua requires $\mathfrak{f}_{mnp}=0$, the gauge group generated by $\mathfrak{f}_{mab}$ must also be $SO(2,1)$. Note also that the compact parts of these two $SO(2,1)$ factors are embedded in the matter and R-symmetry direction, respectively. After taking into account both electric and magnetic components, we find that the gauge group takes the form of
\begin{equation}
SO(2,1)^2_{\textrm{e(m)}}\times SO(2,1)^2_{\textrm{m(e)}}\qquad \textrm{or}\qquad SO(2,2)_+\times SO(2,2)_-\, .
\end{equation}
In this equation, we have used the isomorphism $SO(2,2)_\pm \sim SO(2,1)_\pm \times SO(2,1)_\pm$. We also note here that in this case, there can be three electric (magnetic) and one magnetic (electric) $SO(2,1)$ factors. The number of electric and magnetic factors needs not be equal, but both types of gaugings are required.
\\
\indent We finally consider the most general case of $\mathfrak{f}_{abc}\neq 0$ and $\mathfrak{f}_{mab}\neq 0$. In general, there can be a subspace in which a subset of $\mathfrak{f}_{abc}$ forms a separate compact group $H_{\textrm{c}}$. We will split $\mathfrak{f}_{abc}$ into two parts $\mathfrak{f}_{a'b'c'}$ and $\mathfrak{f}_{a''b''c''}$ with $\mathfrak{f}_{a''mn}=0$. The components $\mathfrak{f}_{a'b'c'}$ together with $\mathfrak{f}_{a'mn}$ form a non-compact group $G_{\textrm{nc}}$ as discussed above while $\mathfrak{f}_{a''b''c''}$ form a separate compact factor $H_{\textrm{c}}$. In addition, the quadratic constraint implies that $\mathfrak{f}_{amn}$ and $\mathfrak{f}_{mab}$ cannot have common indices, so $\mathfrak{f}_{mab}$ again generate an $SO(2,1)$ factor as in the previous case. The gauge group is then given by
\begin{equation}
SO(2,1)_{\textrm{e(m)}}\times SO(2,1)_{\textrm{m(e)}}\times G_{\textrm{nc, e(m)}}\times G'_{\textrm{nc, m(e)}}\times H_{\textrm{c}}\, .
\end{equation}
An example for this type of gauge groups with $n=6$ vector multiplets and $\mathfrak{f}_{a''b''c''}=0$ is given by
\begin{equation}
SO(3,1)_+\times SO(2,1)_+\times SO(2,1)_-
\end{equation}
with $SO(3,1)_+\times SO(2,1)_+$ identified with $G_{\textrm{nc}}\times G'_{\textrm{nc}}$.

%%%%%%%%%%%%%%%%%%%%%%%%%%%%%%%%%%%%%%%%%%%%%%%%%%%%%%%%%%%%%%%
\section{$dS_4$ vacua from different gauge groups}\label{dS4_example}
In this section, we consider gauge groups that lead to $dS_4$ vacua for the case of $n=6$ vector multiplets. These gauge groups have been classified in \cite{de-Roo-Panda-2}. There are nine semi-simple gauge groups that can be embedded in $SO(6,6)$ given by
\begin{eqnarray}
& &SO(2,1)^2_+\times SO(2,1)^2_-,\label{eq:drp-g2}\\
& &SO(3,1)_+\times SO(2,1)_+\times SO(2,1)_-,\label{eq:drp-g3}\\
& & SO(3,1)_+\times SO(3,1)_+ ,\label{eq:drp-g4}\\
& & SO(3)^2_-\times SO(3)^2_+ ,\label{eq:drp-g5}\\
& & SO(3,1)_-\times SO(3)_-\times SO(3)_+,\label{eq:drp-g6}\\
& & SO(3,1)_-\times SO(3,1)_-,\label{eq:drp-g7}\\
& &SO(2,1)^3_+\times SO(3)_+ ,\label{eq:drp-g1}\\
& &SL(3,\mathbb R)_-\times SO(3)_-,\label{eq:drp-g8}\\
& &SU(2,1)_+\times SO(2,1)_+\label{eq:drp-g9}
\end{eqnarray}
in which the extra $SO(2)$ factor in the last two gauge groups has been neglected. In the following analysis, we will explicitly compute the scalar potentials and fermion-shift matrices for these gauge groups and verify that the $dS_4$ vacua satisfy the two sets of conditions given in the previous section. 
\\
\indent However, all these gauge groups have been originally constructed by using the old formulation of \cite{de_Roo_N4_4D}. We need to recast them in the embedding tensor formalism. The first six gauge groups have already been done in \cite{Roest-09} with the corresponding embedding tensors for various factors given by 
\begin{eqnarray}
SO(4)_{\textrm{e(m)}}: && \begin{cases} f_{+123} = \sqrt{2}(g_1-\tilde g_1), \qquad f_{+789} = \sqrt{2}(g_1 + \tilde g_1), \\
f_{-456} = \sqrt{2}(g_2 - \tilde g_2), \qquad f_{-10,11,12} = \sqrt{2}(g_2 + \tilde g_2)\end{cases}\label{eq:so4em} \\\nonumber\\
%%%
SO(3,1)_{\textrm{e(m)}}: && 
\begin{cases} 	f_{+123} = -f_{+783} =f_{+729} = f_{+189} = \frac{1}{\sqrt{2}}(g_1 - \tilde g_1),\\
	 f_{+789} =-f_{+129} = f_{+183} = f_{+723}= \frac{1}{\sqrt{2}}(g_1 + \tilde g_1),\\
 f_{-456} = -f_{10,11,12} = f_{-10,5,12} = f_{-4,11,12} = \frac{1}{\sqrt{2}}(g_2 - \tilde g_2),\\
	f_{-10,11,12} = -f_{-4,5,12} = f_{-4,11,6} = f_{-10,5,6} = \frac{1}{\sqrt{2}}(g_2 + \tilde g_2)\end{cases}\label{eq:so31em} \\\nonumber\\
SO(2,2)_{\textrm{e(m)}}: && 
\begin{cases}
f_{+723} = \frac{1}{\sqrt{2}}(g_1 + \tilde g_1), \qquad f_{+189} = \frac{1}{\sqrt{2}}(g_1 - \tilde g_1), \\
f_{-10,5,6} =\frac{1}{\sqrt{2}}(g_2 + \tilde g_2), \qquad f_{-4,11,12} = \frac{1}{\sqrt{2}}(g_2-\tilde g_2)
 \end{cases}\, . \label{eq:so22em}
\end{eqnarray}
The six gauge groups in \eqref{eq:drp-g2} to \eqref{eq:drp-g7} are obtained by combinations of these $SO(4)$, $SO(3,1)$ and $SO(2,2)$ groups with suitable choices of the coupling constants as shown in table \ref{table:1}.
\\
\indent We also note here that there can be other possible assignments for which simple factor corresponding to electric or magnetic embedding. For example, in $SO(2,2)\times SO(2,2)$ gauge group, we can have only one electric factor of $SO(2,1)$ and three magnetic $SO(2,1)$ factors or vice versa. The embedding tensor in this case is given by
\begin{eqnarray}\label{eq:so22em-3m1e}
& &f_{+723} = \frac{1}{\sqrt{2}}(g_1 + \tilde g_1) , \qquad f_{-189} = \frac{1}{\sqrt{2}}(g_1 - \tilde g_1), \nonumber \\ 
& & f_{-10,5,6} =\frac{1}{\sqrt{2}}(g_2 + \tilde g_2), \qquad f_{-4,11,12} = \frac{1}{\sqrt{2}}(g_2-\tilde g_2).
\end{eqnarray}
The electric-magnetic dual with one magnetic and three electric $SO(2,1)$'s is simply obtained by interchanging $+$ and $-$.
\\
\indent The embedding tensors for the remaining three gauge groups $SO(3)\times SO(2,1)^3$, $SU(2,1)\times SO(2,1)$ and $SL(3, \mathbb R)\times SO(3)$ are obtained as follow.
\begin{itemize}
\item For $SO(3)_+\times SO(2,1)_+^3$, we rewrite it as $SO(3)\times SO(2,1)\times SO(2,1)^2$ with the embedding tensor given by
\begin{equation}\label{eq:SO321}
f_{\alpha\, 569} = g_1, \qquad f_{\alpha\, 10,11,12} = g_2, \qquad f_{\beta\, 127} = \tilde g_1, \qquad f_{\beta\, 348} = \tilde g_2
\end{equation}
with $\alpha=\pm$ and $\beta=\mp$ corresponding to the following electric and magnetic factors $SO(3)_{\textrm{e(m)}}\times SO(2,1)_{\textrm{e(m)}}\times SO(2,1)_{\textrm{m(e)}}^2$
\item For $SU(2,1)_+\times SO(2,1)_+$, we choose the following gauge generators. The $SO(2,1)$ factor is generated by $X_5, X_6$ and $X_{11}$ while the $SU(2,1)$ is generated by $X_{1},\ldots, X_4, X_7,\ldots X_{10}$ with the compact generators being $X_7,\ldots X_{10}$. The associated embedding tensor is given by
\begin{eqnarray}\label{eq:SU2121}
& &f_{\alpha\, 129}  = f_{\alpha\,138}=f_{\alpha\,147} =  f_{\alpha\,248} =-g_1,\qquad f_{\alpha\,237}=f_{\alpha\,349} = g_1,\nonumber \\ & &f_{\alpha\,789} = 2g_1, \qquad f_{\alpha\,1,2,10} = f_{\alpha\,3,4,10}=-\sqrt{3}g_1,\qquad f_{\beta\,5,6,11} = g_2
\end{eqnarray}
with $\alpha = \pm$ and $\beta =\mp$ corresponding to $SU(2,1)_{\textrm{e(m)}}\times SO(2,1)_{\textrm{m(e)}}$.
\item For $SL(3,\mathbb R)_-\times SO(3)_-$, we choose the generators for $SO(3)$ to be $X_4, X_5$ and $X_6$ while $SL(3,\mathbb R)$ is generated by the compact $X_1, X_2, X_3$ and non-compact $X_7,\ldots, X_{11}$ generators. Non-vanishing components of the embedding tensor are given by
\begin{eqnarray}\label{eq:SL3SO3}
& &f_{\alpha\,123} = f_{\alpha\,1,9,10}=f_{\alpha\,279}=-f_{\alpha\,2,8,10}=f_{\alpha\,3,7,10}=f_{\alpha\,3,8,9}=-g_1,\nonumber \\ 
& &f_{\alpha\,178}=2g_1,\qquad f_{\alpha\,2,10,11}=f_{\alpha\,3,9,11} =\sqrt{3}g_1, \qquad f_{\beta\, 456} = g_2
\end{eqnarray}
with $\alpha = \pm$ and $\beta =\mp$ corresponding to $SL(3,\mathbb R)_{\textrm{e(m)}}\times SO(3)_{\textrm{m(e)}}$.
\end{itemize}
%%%%%%%%%%%%%%%%%%%%%%%%%%%%%%%
\begin{table}[!htbp]
\centering
\begin{tabular}{|l|l|l|}
\hline
 Gauge groups in \cite{de-Roo-Panda-2} & Gauge groups in \cite{Roest-09} & Conditions\\
\hline
%&&\\
 $SO(3)^2_-\times SO(3)^2_+$ &  $SO(4)_{\textrm{e}} \times SO(4)_{\textrm{m}}$ & $g_1, \tilde g_1, g_2, \tilde g_2 \neq 0$\\
 %&&\\
 \hline
 $SO(3,1)_+\times SO(3,1)_+$ & $SO(3,1)_{\textrm{e}}\times SO(3,1)_{\textrm{m}}$ & $\begin{array}{l}\tilde g_1 = g_1, \tilde g_2 = g_2,\\ g_1, g_2 \neq 0\end{array}$\\
 %&&\\
 \hline
 $SO(3,1)_-\times SO(3,1)_-$ & $SO(3,1)_{\textrm{e}}\times SO(3,1)_{\textrm{m}}$ & $\begin{array}{l}\tilde g_1 = -g_1, \tilde g_2 = -g_2,\\ g_1, g_2 \neq 0 \end{array}$\\
 %&&\\
 \hline
 $SO(3,1)_-\times SO(3)_-\times SO(3)_+$ & $\begin{array}{l}SO(3,1)_{\textrm{m}}\times SO(4)_{\textrm{e}}\\\\SO(3,1)_{\textrm{e}}\times SO(4)_m\end{array}$& $\begin{array}{l}\tilde g_2 = -g_2, \\g_2, g_1, \tilde g_1 \neq 0\\ \\\tilde g_1 = -g_1,\\ g_1, g_2, \tilde g_2 \neq 0\end{array}$ \\
 %&&\\
 \hline
 $SO(2,1)^2_+\times SO(2,1)^2_-$ & $SO(2,2)_{\textrm{e}} \times SO(2,2)_{\textrm{m}}$ &  $g_1, \tilde g_1, g_2, \tilde g_2 \neq 0$\\
 %&&\\
 \hline
 $SO(3,1)_+\times SO(2,1)_+\times SO(2,1)_-$ & $\begin{array}{l}SO(3,1)_{\textrm{m}}\times SO(2,2)_{\textrm{e}}\\ \\SO(3,1)_{\textrm{e}}\times SO(2,2)_{\textrm{m}}\end{array}$ & $\begin{array}{l}\tilde g_2 = g_2,\\ g_2, g_1,\tilde g_1 \neq 0 \\\\ \tilde g_1 = g_1,\\ g_1, g_2, \tilde g_2 \neq 0\end{array}$ \\
\hline
\end{tabular}
\caption{The six gauge groups giving rise to $dS_4$ vacua as given in \cite{de-Roo-Panda-2}. The embedding tensors for these gauge groups are obtained by imposing some relations between the coupling constants as shown in the last column.}\label{table:1}
\end{table}
\indent We now compute the fermion-shift matrices and scalar potential. To do an explicit computation, we will work with the coset representative $\mc{V}_\alpha$ of the form  
\begin{eqnarray}
\mc V_\alpha = e^{\phi/2} \begin{pmatrix}\chi - i e^{-\phi} \\ 1\end{pmatrix}\, .
\end{eqnarray}
Since all known $dS_4$ vacua are found only with vanishing scalars from vector multiplets, we will give the scalar potential only for non-vanishing dilaton $\phi$ and axion $\chi$ to simplify the results. 
\\
\indent With all $SO(6,6)/SO(6)\times SO(6)$ scalars set to zero, we simply have 
\begin{equation}
\mc{V}=\mathbf{I}_{12}
\end{equation}
and ${\mc{V}_M}^{ij}=\frac{1}{2}\Gamma^{ij}_m{\mc{V}_M}^m$. Note that an extra factor of $\frac{1}{2}$ is added for consistency with the normalization used in \cite{N4_gauged_SUGRA} for the following $SO(6,n)$ identity 
\begin{equation}
\eta_{MN}=-\frac{1}{2}\epsilon_{ijkl}{\mc{V}_M}^{ij}{\mc{V}_N}^{kl}+{\mc{V}_M}^a{\mc{V}_N}^a\, .
\end{equation}
An explicit form of $SO(6)$ gamma matrices is given in the appendix. We are now in a position to consider each gauge group in detail. We emphasize that all of the critical points considered here are already known. Our main aim is to verify that they satisfy the conditions introduced in the previous section.  For convenience, we collect the conditions for the existence of $dS_4$ vacua given in sections \ref{dS4_1} and \ref{dS4_2} here 
\begin{eqnarray}
&&\langle A_1^{ij}\rangle=\langle {A_{2ai}}^{j}\rangle=0, \qquad \langle A_2^{ij}A^*_{2kj}\rangle =\frac{9}{4} V_0\delta^i_k, \label{eq:dS4set1}\\
 && \langle A_1^{ij}\rangle=\langle A_{2}^{ij}\rangle=0, \qquad \langle {A_{2ak}}^{j}{A^*_{2aj}}^i\rangle =\frac{1}{2} V_0\delta^i_k\, .\label{eq:dS4set2}
\end{eqnarray}
We will also refer to $dS_4$ vacua as the first and second type $dS_4$ if they satisfy \eqref{eq:dS4set1} and \eqref{eq:dS4set2}, respectively.

\subsection{$SO(3)^2_+\times SO(3)_-^2$}
This case corresponds to the gauging of $SO(4)_{\textrm{e}}\times SO(4)_{\textrm{m}}$ group. The embedding tensor for this gauge group is given in \eqref{eq:so4em}. The scalar potential is found to be
\begin{equation}
V=-e^{\phi } \left[g_1^2-2 g_1 \tilde g_1+\chi ^2 (g_2-\tilde g_2)^2+\tilde g_1^2\right]+4 (g_1-\tilde g_1) (g_2-\tilde g_2)-e^{-\phi } (g_2-\tilde g_2)^2
\end{equation}
with the following critical point
\begin{eqnarray}
\chi = 0, \qquad \phi = \ln \left[\pm \frac{g_2 - \tilde g_2}{g_1 - \tilde g_1}\right].
\end{eqnarray}
To bring this critical point to the origin $\chi = \phi =0$, we have two possibilities:
\begin{itemize}
	\item Setting $\tilde g_2 -g_2= g_1 -\tilde g_1$ leads to an $AdS_4$ critical point which is the trivial critical point of the same gauge group reported in \cite{4D_N4_flows} with $V_0 = -6(g_1 -\tilde g_1)^2$. As expected, this critical point satisfies the $AdS_4$ conditions given in \cite{AdS4_N4_Jan} 
\begin{eqnarray}\label{eq:so4so4AdS4}
\langle A_2^{ij}\rangle=\langle {A_{2ai}}^j\rangle=0, \qquad \langle A_1^{ij}A^*_{1kj}\rangle =  -\frac{4}{3} V_0 \delta^i_k\, .
\end{eqnarray}
%%%
\item Another possibility is to set $\tilde g_2 -g_2= -(g_1 - \tilde g_1)$ which leads to a $dS_4$ critical point with $V_0 = 2(g_1-\tilde g_1)^2$ and satisfying the conditions in \eqref{eq:dS4set1}
\begin{eqnarray}\label{eq:so4so4dS4}
 \langle A_1^{ij}\rangle=\langle {A_{2ai}}^{j}\rangle=0, \qquad \langle A_2^{ij}A^*_{2kj}\rangle = V_0\delta^i_k\, .
\end{eqnarray}
\end{itemize}

\subsection{$SO(3,1)_\pm\times SO(3,1)_\pm$}
The embedding tensor for this gauge group is given in \eqref{eq:so31em} with the scalar potential given by
\begin{eqnarray}
V&=&\frac{1}{2} e^{-\phi } \left[e^{2 \phi } \left(g_1^2+4 g_1 \tilde g_1+\chi ^2 \left(g_2^2+4 g_2 \tilde g_2+\tilde g_2^2\right)+\tilde g_1^2\right)\right.\nonumber\\
&&\left.+2 e^{\phi } (g_1-\tilde g_1) (g_2-\tilde g_2)+g_2^2+4 g_2 \tilde g_2+\tilde g_2^2\right].
\end{eqnarray}
There is a critical point at
\begin{eqnarray}\label{eq:3131cp}
\chi=0,\qquad \phi = \ln\left[\pm\frac{g_2^2 + 4g_2 \tilde g_2 + \tilde g_2^2}{g_1^2 + 4 g_1 \tilde g_1 + \tilde g_1^2}\right].
\end{eqnarray}
We now separately consider two types of gauge groups.
\begin{itemize}
	\item For $SO(3,1)_+\times SO(3,1)_+$, we choose $\tilde g_1 = g_1$ and $\tilde g_2 = g_2$ which eliminate the following components of the embedding tensor
\begin{eqnarray}
&& f_{+123} = -f_{+783} =f_{+729} = f_{+189}=0, \nonumber\\
&& f_{-456} = -f_{10,11,12} = f_{-10,5,12} = f_{-4,11,12} =0\, .
\end{eqnarray}
Accordingly, the $SO(3)$ subgroups of both $SO(3,1)$ factors are embedded along the matter-multiplet directions $M =7,8,9$ and $M=10,11,12$. Setting $\tilde g_1 = g_1$ and $\tilde g_2 = g_2$, we can rewrite the critical point \eqref{eq:3131cp} as
\begin{eqnarray}
\chi = 0, \qquad \phi = \ln\left[\pm\frac{g_2}{g_1}\right]
\end{eqnarray}
To bring this critical point to the values $\chi= \phi =0$, we set $g_2 = \pm g_1$, and both of these choices lead to the same $dS_4$ critical point with $V_0 =6g_2^2$ and satisfying \eqref{eq:dS4set2}
\begin{eqnarray}\label{eq:so31so31dS4-2}
\langle A_1^{ij}\rangle =\langle A_2^{ij}\rangle =0, \qquad \langle {A_{2ai}}^{k}{A^*_{2ak}}^j\rangle = 3g_2^2\delta^j_i\, .
\end{eqnarray}
This $dS_4$ vacuum is then of the second type. There is no $AdS_4$ vacuum in this case since the existence of $AdS_4$ requires the embedding of $SO(3)\times SO(3)$ along the R-symmetry directions.

\item For $SO(3,1)_-\times SO(3,1)_-$, we set $\tilde g_1 = -g_1$ and $\tilde g_2 = -g_2$ which give
\begin{eqnarray}
 f_{+789} =-f_{+129} = f_{+183} = f_{+723}= 0,\nonumber\\
 f_{-10,11,12} = -f_{-4,5,12} = f_{-4,11,6} = f_{-10,5,6} =0\, .\nonumber
\end{eqnarray}
The $SO(3)$ subgroups of both $SO(3,1)$ factors are now embedded along the R-symmetry directions $M=1,2,3$ and $M=4,5,6$. With this choice of the coupling constants, the critical point \eqref{eq:3131cp} becomes
\begin{eqnarray}
\chi = 0, \qquad \phi = \ln\left[\pm\frac{g_2}{g_1}\right].
\end{eqnarray}
In this case, however, setting $g_2 = \pm g_1$ leads to two different critical points.
\begin{itemize}
	\item Setting $g_2=g_1$ leads to a $dS_4$ critical point satisfying \eqref{eq:dS4set1}
\begin{eqnarray}\label{eq:so31so31dS4}
 V_0 = 2g_2^2,  \qquad \langle A_1^{ij}\rangle=\langle {A_{2ai}}^{j}\rangle=0, \qquad \langle A_2^{ij}A^*_{2kj}\rangle = \frac{9}{4}V_0\delta^i_k \, .
\end{eqnarray}
\item The choice $g_2 = -g_1$ gives an $AdS_4$ critical point with $V_0 = -6g_2^2$ and satisfying 
\begin{eqnarray}\label{eq:so31so31AdS4}
\langle A_2^{ij}\rangle=\langle {A_{2ai}}^{j}\rangle=0, \qquad \langle A_1^{ij}A^*_{1kj}\rangle = -\frac{3}{4}V_0\delta^i_k\, .
\end{eqnarray}
\end{itemize}
In this case, the $dS_4$ vacuum is of the first type. It should be noted that, as in the $SO(3)^2_+\times SO(3)^2_-$ gauge group, the $SO(3,1)_-\times SO(3,1)_-$ gauge group gives two types of vacua with opposite ratios of the coupling constants. 
\end{itemize}
It is useful to note that, gauge groups of the form $SO(3,1)_\pm\times SO(3,1)_\mp$, obtained by setting $\tilde{g}_1=\pm g_1$ and $\tilde{g}_2=\mp g_2$, lead to a Minkowski vacuum. 

\subsection{$SO(2,1)^2\times SO(2,1)^2$}
In this case, the gauge group is given by $SO(2,2)\times SO(2,2)\sim SO(2,1)^4$, and there are two possible gaugings to consider depending on the asignment of electric or magnetic gaugings to each $SO(2,1)$ factor. One gauging is described by $SO(2,2)_{\textrm{e}}\times SO(2,2)_{\textrm{m}}\sim SO(2,1)_{\textrm{e}}\times SO(2,1)_{\textrm{e}}\times SO(2,1)_{\textrm{m}}\times SO(2,1)_{\textrm{m}}$ with the embedding tensor given in \eqref{eq:so22em}. The other one is $SO(2,1)_{\textrm{e}}\times SO(2,1)_{\textrm{m}}\times SO(2,1)_{\textrm{m}}\times SO(2,1)_{\textrm{m}}$ with the embedding tensor given in \eqref{eq:so22em-3m1e} and its electric-magnetic dual $SO(2,1)_{\textrm{e}}\times SO(2,1)_{\textrm{e}}\times SO(2,1)_{\textrm{e}}\times SO(2,1)_{\textrm{m}}$.
\\
\indent It turns out that all of these gaugings give rise to the same scalar potential of the form
\begin{eqnarray}
V=\frac{1}{4} e^{-\phi } \left[e^{2 \phi } \left[(g_1+\tilde g_1)^2+\chi ^2 (g_2+\tilde g_2)^2\right]+(g_2+\tilde g_2)^2\right]
\end{eqnarray}
with the following critical point
\begin{eqnarray}\label{eq:2121cp}
\chi = 0, \qquad \phi = \ln\left[\pm \frac{g_2 + \tilde g_2}{g_1+\tilde g_1}\right].
\end{eqnarray}
The critical point can be shifted to the origin $\chi = 0$ and $\phi =0$ by setting $\tilde g_2 +g_2= \pm (g_1 + \tilde g_1)$. Both choices lead to the same $dS_4$ critical point with $V_0 = \frac{1}{2}(g_1 + \tilde g_1)^2$ and satisfying \eqref{eq:dS4set2}
\begin{eqnarray}\label{eq:so22so22dS4-2}
\langle A_1^{ij}\rangle=\langle A_{2}^{ij}\rangle=0, \qquad \langle {A_{2ai}}^{k}{A^*_{2ak}}^j\rangle = \frac{1}{2}V_0\delta^j_i\, .
\end{eqnarray}

\subsection{$SO(3)_+\times SO(3)_-\times SO(3,1)_-$}
This case corresponds to $SO(4)\times SO(3,1)$ gauge group with two possible gaugings $SO(4)_{\textrm{e(m)}}\times SO(3,1)_{\textrm{m(e)}}$. The embedding tensors are given by
	\begin{eqnarray}
	\textrm{I}:\qquad &&SO(3)_{\textrm{e}+}\times SO(3)_{\textrm{e}-}\times SO(3,1)_{\textrm{m}-}:\nonumber\\
	 && f_{+123} = g_1, \hspace{4mm} f_{+789} = \tilde g_1\nonumber\\
	 && f_{-456} = -f_{-10,11,6} = f_{-10,5,12} = f_{-4,11,12} = g_2, \label{eq:431em}\\\nonumber\\
	\textrm{II}:\qquad &&SO(3,1)_{\textrm{e}-}\times SO(3)_{\textrm{m}-}\times SO(3)_{\textrm{m}+}:\nonumber\\
	&& f_{+123} = -f_{+783} = f_{+729}=f_{+189} = g_1\nonumber\\
	 && f_{-456} = g_2, \qquad f_{10,11,12} = \tilde g_2\, .\label{eq:314em}
	\end{eqnarray}
Note that the embedding tensors for $SO(3,1)$'s in both cases are obtained from \eqref{eq:so31em} by setting $\tilde g_2 = -g_2$ and $\tilde g_1 = -g_1$, respectively. The two gaugings lead to the same scalar potential given by 
	\begin{eqnarray}
V = 2 g_1 g_2 -\frac{1}{2}e^{-\phi}g_2^2- \frac{1}{2}e^\phi (g_1^2 + g_2^2\chi^2).
	\end{eqnarray}
	This potential admits a critical point at
	\begin{eqnarray}
	\chi = 0, \qquad\phi = \ln\left[\pm \frac{g_2}{g_1}\right]
	\end{eqnarray}
which can be shifted to the origin $\chi =\phi =0$ by setting $g_2=\pm g_1$. We now look at these two choices.
\begin{itemize}
	\item The case of $g_2 = g_1$ leads to a $dS_4$ solution with $V_0 = g_1^2$ and satisfies \eqref{eq:dS4set1}
	\begin{eqnarray}\label{eq:so4so31dS4}
\langle A_1^{ij}\rangle = \langle {A_{2ai}}^{j}\rangle =0, \qquad \langle A_2^{ij}A_{2kj}^*\rangle = \frac{9}{4}V_0\delta^i_k\, .
	\end{eqnarray}
\item For $g_2 = -g_1$, the critical point is an $AdS_4$ vacuum with $V_0 = -3g_1^2$.
	\end{itemize}
	
\subsection{$SO(3,1)_+\times SO(2,1)_+\times SO(2,1)_-$}
This case corresponds to $SO(2,2)\times SO(3,1)$ gauge group with two possible gaugings, $SO(2,2)\sim SO(2,1)\times SO(2,1)$ and $SO(3,1)$ factors being electric and magnetic or vice versa.
	\begin{eqnarray}
	\textrm{I}: \qquad & & SO(2,2)_\textrm{e}\times SO(3,1)_\textrm{m}: \nonumber\\
	&&f_{+723} =\frac{1}{\sqrt{2}}(g_1 + \tilde g_1), \qquad f_{+189} = \frac{1}{\sqrt{2}}(g_1 -\tilde g_1), \nonumber\\
	&&  f_{-456} = -f_{10,11,12} = f_{-10,5,12} = f_{-4,11,12} = \frac{1}{\sqrt{2}}(g_2 - \tilde g_2),\nonumber\\
	&&f_{-10,11,12} = -f_{-4,5,12} = f_{-4,11,6} = f_{-10,5,6} = \frac{1}{\sqrt{2}}(g_2 + \tilde g_2).\\
	\textrm{II}: \qquad & & SO(2,2)_\textrm{m} \times SO(3,1)_\textrm{e}: \nonumber\\
	&&f_{+123} = -f_{+783} =f_{+729} = f_{+189} = \frac{1}{\sqrt{2}}(g_1 - \tilde g_1),\nonumber\\
	&& f_{+789} =-f_{+129} = f_{+183} = f_{+723}= \frac{1}{\sqrt{2}}(g_1 + \tilde g_1),\nonumber\\
	&& f_{-10,5,6} = \frac{1}{\sqrt{2}}(g_2 + \tilde g_2), \qquad f_{-4,11,12} = \frac{1}{\sqrt{2}}(g_2 - \tilde g_2).
	\end{eqnarray}
	In order to have $SO(3,1)_+$, we will set $\tilde{g}_2=g_2$ and $\tilde{g}_1=g_1$, respectively. These two gaugings lead to the scalar potentials
	\begin{eqnarray}
	V_{\textrm{I}} &=& \frac{1}{4} e^{\phi } \left[g_1^2+2 g_1 \tilde g_1+2 \chi ^2 \left(g_2^2+4 g_2 \tilde g_2+\tilde g_2^2\right)+\tilde g_1^2\right]+\frac{1}{2} e^{-\phi } \left(g_2^2+4 g_2 \tilde g_2+\tilde g_2^2\right)\nonumber\\
	& &\\
	V_{\textrm{II}} &=& \frac{1}{4} e^{-\phi } \left[e^{2 \phi } \left[2 g_1^2+8 g_1 \tilde g_1+\chi ^2 (g_2+\tilde g_2)^2+2 \tilde g_1^2\right]+(g_2+\tilde g_2)^2\right]
	\end{eqnarray}
	We now look at critical points of these potentials.
	\begin{itemize}
	\item The critical point of $V_{\textrm{I}}$ is given by
	\begin{eqnarray}
	\chi = 0, \qquad \phi = \ln\left[\pm \frac{\sqrt{2(g_2^2+4g_2 \tilde g_2 + \tilde g_2^2)}}{(g_1+\tilde g_1)}\right]
	\end{eqnarray}
	which can be brought to the origin by choosing
	\begin{eqnarray}
	\tilde g_1 = -g_1 \pm 2\sqrt{3} g_2
	\end{eqnarray}
	after setting $g_2 = \tilde g_2$. Both choices lead to the same $dS_4$ vacuum with $V_0 =6g_2^2$ and satisfying \eqref{eq:dS4set2}.

	\item For $V_{\textrm{II}}$, we find the following critical point
	\begin{eqnarray}
	\chi = 0, \qquad \phi = \ln\left[\pm \frac{(g_2 + \tilde g_2)}{\sqrt{2(g_1^2 + 4 g_1 \tilde g_t + \tilde g_1^2)}}\right].
	\end{eqnarray}
After setting $\tilde g_1 = g_1$, this critical point can be brought to the origin by choosing
	\begin{eqnarray}
	\tilde g_2 = -g_2 \pm 2\sqrt{3}g_1\, .
	\end{eqnarray}
Both sign choices again give the same $dS_4$ critical point which satisfies \eqref{eq:dS4set2} and $V_0 = 6g_1^2$.
\end{itemize}

\subsection{$SO(3)_+\times SO(2,1)_+^3$}
The embedding tensor for this gauge group is given in \eqref{eq:SO321}. The two possible gaugings $SO(3)_{\textrm{e(m)}}\times SO(2,1)_{\textrm{e(m)}}\times SO(2,1)^2_{\textrm{m(e)}}$ respectively give the following scalar potentials and critical points
\begin{eqnarray}
V_{\textrm{I}}&=&
\frac{1}{2} e^{-\phi } \left[g_1^2 \left(\chi ^2 e^{2 \phi }+1\right)+e^{2 \phi } \left(\tilde g_1^2+\tilde g_2^2\right)\right], \nonumber \\
\chi &=& 0,\qquad \phi = \ln\left[\pm \frac{g_1}{\sqrt{\tilde g_1^2 + \tilde g_2^2}}\right]
\end{eqnarray}
and 
\begin{eqnarray}
V_{\textrm{II}}&=&
\frac{1}{2} e^{-\phi } \left[e^{2 \phi } \left(g_1^2+\chi ^2 \left(\tilde g_1^2+\tilde g_2^2\right)\right)+\tilde g_1^2+\tilde g_2^2\right],\nonumber \\
\chi &=& 0,\qquad  \phi = \ln\left[\pm \frac{\sqrt{\tilde g^2_1 + \tilde g^2_2}}{g_1}\right].
\end{eqnarray}
These critical points can be shifted to the origin by setting $g_1=\pm \sqrt{\tilde g^2_1+ \tilde g^2_2}$, leading to the same $dS_4$ solution with $V_0 = g_1^2$ and satisfying \eqref{eq:dS4set2}.

\subsection{$SU(2,1)_+\times SO(2,1)_+$}
The embedding tensor for this gauge group is given in \eqref{eq:SU2121}. This gauge group can be embedded either as $SU(2,1)_{\textrm{e}}\times SO(2,1)_{\textrm{m}}$ or $SU(2,1)_{\textrm{m}}\times SO(2,1)_{\textrm{e}}$. The scalar potentials and critical points for these two gaugings are given by
\begin{eqnarray}
\textrm{I}:\qquad V_{\textrm{I}} &=&  
\frac{1}{2}  \left[e^{ \phi } \left(12 g_1^2+g_2^2 \chi ^2\right)+e^{-\phi }g_2^2\right],\nonumber \\
\chi &=&0, \qquad \phi = \ln\left[\pm \frac{g_2}{2\sqrt{3}g_1}\right],\\
\textrm{II}:\qquad V_{\textrm{II}} &=& \frac{1}{2} e^{\phi } \left(12 g_1^2 \chi ^2+g_2^2\right)+6 g_1^2 e^{-\phi },\nonumber \\
\chi &=&0, \qquad \phi = \ln\left[\pm \frac{2\sqrt{3}g_1}{g_2}\right].
\end{eqnarray}
Choosing $g_2 = \pm 2\sqrt{3}g_1$ leads to a $dS_4$ solution satisfying \eqref{eq:dS4set2} with $V_0 = 12 g_1^2$.

\subsection{$SL(3,\mathbb R)_-\times SO(3)_-$ }
The embedding tensor for this gauge group is given in \eqref{eq:SL3SO3}. In this gauge group, both $AdS_4$ and $dS_4$ solutions are possible, and, unlike the previous cases, the choice of which gauge group factor is electric or magnetic affects the resulting solutions. We will consider each choice separately.
\\
\indent For $SL(3,\mathbb R)_\textrm{e}\times SO(3)_\textrm{m}$ embedding, the scalar potential is given by
\begin{equation}
V = -\frac{1}{2}  \left[e^{\phi } \left(g_1^2+g_2^2 \chi ^2\right) +e^{-\phi }g_2^2+4 g_1 g_2\right]
\end{equation}
with a critical point at
\begin{equation}
\chi = 0, \qquad \phi = \ln\left[\pm \frac{g_2}{g_1}\right].
\end{equation}
Choosing $g_2=g_1$ and $g_2=-g_1$ leads to $AdS_4$ and $dS_4$ vacua with $V_{0}=-3g_1^2$ and $V_0=g_1^2$, respectively. The $dS_4$ vacuum satisfies the relations given in \eqref{eq:dS4set1}. 
\\
\indent For $SL(3,\mathbb R)_{\textrm{m}}\times SO(3)_{\textrm{e}}$ embedding, we find a similar structure with the scalar potential and critical point given by
\begin{equation}
V=-\frac{1}{2} \left[e^{\phi } \left(g_1^2 \chi ^2+g_2^2\right)+g_1^2 e^{-\phi }-4 g_1 g_2\right]
\end{equation}
and 
\begin{equation}
\chi = 0, \qquad\phi = \ln\left[\pm \frac{g_1}{g_2}\right].
\end{equation}
Choosing $g_2=-g_1$ and $g_2=g_1$ leads to $AdS_4$ and $dS_4$ vacua with $V_{0}=-3g_1^2$ and $V_0=g_1^2$, respectively.

%%%%%%%%%%%%%%%%%%%%%%%%%%%%%%%%%%%%%%%%%%%%%%%%%%%%%%%%%%%
\section{Relations between gaugings with $dS_4$ and $dS_5$ vacua}\label{dS4_dS5}
In this section, we give some relations between gaugings of $N=4$ gauged supergravities in four and five dimensions with de Sitter vacua. In general, a circle reduction of $N=4$ five-dimensional theory gives rise to four-dimensional theory with the same number of supersymmetries. As pointed out in \cite{N4_gauged_SUGRA}, the relations between the embedding tensors in four and five dimensions can be obtained from an analysis of group structures. We will follow this procedure in relating four- and five-dimensional gaugings with de Sitter vacua.
\\
\indent A five-dimensional supergravity theory with $\hat{n}$ vector multiplets gives, via a reduction on $S^1$, a four-dimensional theory with $n=\hat{n}+1$ vector multiplets. The global or duality symmetries in these two theories are given by $\hat{G}=SO(1,1)\times SO(5,\hat{n})$ and $G=SL(2)\times SO(6,\hat{n}+1)$, respectively. Accordingly, it is possible that gaugings in five dimensions can be encoded in those in four dimensions since $\hat{G}\subset G$.
\\
\indent Recall that components of the embedding tensor consistent with $N=4$ supersymmetry in five dimensions are given by $\hat \xi_M$, $\hat \xi_{MN}=\hat \xi_{[MN]}$ and $\hat f_{MNP}=\hat f_{[MNP]}$, for more detail see \cite{N4_gauged_SUGRA}. To identify these components with those in four dimensions $\xi_{\alpha M}$ and $f_{\alpha MNP}$, we first consider the decomposition of a representation $(\mathbf{2},\mathbf{n}+\mathbf{7})$ of $SL(2)\times SO(6,n+1)$ under its $SO(1,1)_B\times SO(1,1)_A\times SO(5,n)$ subgroup as follow
\begin{equation}
(\mathbf{2},\mathbf{7}+\mathbf{n})\rightarrow (\mathbf{2},(\mathbf{n}+\mathbf{5})_0)+(\mathbf{2},\mathbf{1}_{\frac{1}{2}})+(\mathbf{2},\mathbf{1}_{-\frac{1}{2}})
\end{equation}
for $SO(6,n+1)\rightarrow SO(1,1)_A\times SO(5,n)$ and 
\begin{equation}
(\mathbf{2},\mathbf{7}+\mathbf{n})\rightarrow (\mathbf{7}+\mathbf{n})_{\frac{1}{2}}+(\mathbf{7}+\mathbf{n})_{-\frac{1}{2}}
\end{equation}
for $SL(2)\rightarrow SO(1,1)_B$. The subscript denotes $SO(1,1)$ charges. These decompositions suggest the split of indices $M=(\hat{M},\oplus, \ominus)$ and $\alpha=(+,-)$. Accordingly, the four-dimensional vector fields $A^{\alpha M}_\mu$ are split into
\begin{displaymath}
A^{\alpha M}_\mu =(A^{\hat{M}+}_\mu,A^{\hat{M}-}_\mu,A^{\oplus +}_\mu,A^{\oplus-}_\mu,A^{\ominus+}_\mu,A^{\ominus -}_\mu).   
\end{displaymath}
\indent The $SO(1,1)$ factor in $\hat{G}$ is identified with the diagonal subgroup of $SO(1,1)_A\times SO(1,1)_B$. Generators of $SO(1,1)\times SO(5,n)$ are denoted by $\hat{t}_{\hat{0}}$ and $\hat{t}_{\hat{M}\hat{N}}$ and given in terms of $SL(2)\times SO(6,n+1)$ generators $(t_{\alpha\beta},t_{MN})$ as follow    
\begin{equation}      
 \hat{t}_{\hat{0}} = t_{+-} + t_{\ominus\oplus}\qquad \textrm{and}\qquad \hat{t}_{\hat{M}\hat{N}} = t_{\hat{M}\hat{N}}\, .     
\end{equation}
The five dimensonal vector fields $(\hat{A}^{\hat{0}}_\mu,\hat{A}^{\hat{M}}_\mu)$ are given by 
\begin{equation}      
 \hat{A}^{\hat{0}}_{\mu} = A^{\ominus -}_\mu \qquad \textrm{and}\qquad \hat{A}^{\hat{M}}_\mu = A^{\hat{M}+}_\mu\, .     
\end{equation}
The vector fields $A^{\hat{M}-}_\mu$ and $A_\mu^{\oplus +}$ are the magnetic dual of $A^{\hat{M}+}_\mu$ and $A^{\ominus -}_\mu$ which arise from the two-form fields in five dimensions. $A^{\oplus -}$ and $A^{\ominus +}$ are uncharged under the $SO(1,1)$ duality group and are the Kaluza-Klein vector coming from the five-dimensional metric and its dual.
\\
\indent By comparing the gauge covariant derivatives in four and five dimensions, we have the following identification of various components of the embedding tensors
\begin{equation}
\xi_{+\hat{M}}=\hat{\xi}_{\hat{M}},\qquad f_{+\hat{M}\oplus \ominus}=\frac{1}{2} \hat{\xi}_{\hat{M}},\qquad f_{-\ominus \hat{M}\hat{N}}=\hat{\xi}_{\hat{M}\hat{N}},\qquad f_{+\hat{M}\hat{N}\hat{P}}=\hat{f}_{\hat{M}\hat{N}\hat{P}}
\end{equation} 
with all the remaining components set to zero in a simple circle reduction. In our analysis of de Sitter vacua, we have $\xi_{\hat{M}}=0$ and $\xi_{\alpha M}=0$, so the relevant relations are given by
\begin{equation}
f_{-\ominus \hat{M}\hat{N}} = \hat{\xi}_{\hat{M}\hat{N}}\qquad \textrm{and} \qquad f_{+\hat{M}\hat{N}\hat{P}} = \hat{f}_{\hat{M}\hat{N}\hat{P}}\, .\label{eq:4d5d}
\end{equation}
\indent In the following analysis, we will give the connections between four- and five-dimensional gaugings that lead to de Sitter vacua. The five-dimensional gaugings have been classified in \cite{dS5}. We will mainly work with $\hat{n}=5$ except for the last example in which $\hat{n}=7$. The latter leads to a new gauge group with a new $dS_4$ vacuum that has not been considered before since all the previous works have been done only for $n=\hat{n}+1=6$. Another point to be noted is that, to apply the above decomposition and identification, we need to relabel some indices and coupling constants and interchange gauge generators while keep track of the R-symmetry and matter multiplet directions. The modifications do not qualitatively change the structure of the gauge groups, scalar potentials and the critical points. 
\\
\indent Finally, we will divide the discussion into two parts since there are two classes of gauge groups in four dimensions that lead to de Sitter vacua. As we will see, the gauge groups with $dS_4$ vacua of the first type lead to gauge groups with only $AdS_5$ vacua in five dimensions in agreement with the absence of the five-dimensional analogue for $dS_4$ vacua of the first type. On the other hand, gauge groups giving rise to $dS_4$ vacua of the second type do lead to five-dimensional gauge groups with $dS_5$ vacua. This fact could possibly be inferred from the similar stucture of four- and five-dimensional gauge groups namely a product of non-compact factors. Before discussing relations between these gauge groups in detail, we first give a summary of four- and five-dimensional gauge groups that are related to each other in table \ref{table2}.
\begin{table}[!htbp]
\centering
\begin{tabular}{|c|c|c|}
\hline& & \\
$\#_\text{5D}$ & 5D gauge groups  &4D gauge groups \\& & \\
\hline& & \\
 1&   $U(1)\times SU(2)\times SU(2)$  &$SO(3)^2_+\times SO(3)^2_-$  \\& & \\
 2&  $U(1)\times SO(3,1)$ &  $SO(3)_-\times SO(3)_+\times SO(3,1)_-$  \\& & \\
 3 & $U(1)\times SL(3,\mathbb R)$ & $SO(3)_-\times SL(3,\mathbb R)_-$ \\& & \\\hline
  & &  \\
  4& $SO(1,1)\times SU(2,1)$ &$SO(2,1)_+\times SU(2,1)_+$  \\& & \\
  5& $SO(1,1)\times SO(2,1)$ &$SO(2,1)_+\times SO(2,1)_+$  \\& & \\
 6 & $SO(1,1)^{(2)}_\textrm{diag}\times SO(2,1)$  &$SO(3,1)_+\times SO(2,1)_+\times SO(2,1)_-$  \\& & \\
  7& $SO(1,1)^{(3)}_\textrm{diag}\times SO(2,1)$ & $SU(2,1)_+\times SO(2,1)_+$  \\& & \\
  8& $SO(1,1)\times SO(2,1)^2$ & $SO(2,1)^2_+\times SO(2,1)_-^2$ \\& & \\
  9& $SO(1,1)\times SO(3,1)$ &  $SO(3,1)_+\times SO(2,1)_+\times SO(2,1)_-$ \\& & \\
  10& $SO(1,1)^{(2)}_\text{diag}\times SO(3,1)$ &$SO(3,1)_+\times SO(3,1)_+$  \\& & \\
  11& $SO(1,1)\times SO(4,1)$ &  $SO(2,1)_+\times SO(4,1)_+$  \\
  \hline
\end{tabular}
\caption{The identification between five- and four-dimensional gaugings. Gauge groups with $\#1,2,3$ in five dimensions admit only $AdS_5$ vacua and are identified with four-dimensional gaugings with $dS_4$ vacua satisfying the conditions \eqref{eq:dS4set1}. Gauge groups with $\# 4,\ldots, 11$ give $dS_5$ vacua and are identified with four-dimensional gaugings that lead to $dS_4$ vacua satisfying the conditions \eqref{eq:dS4set2}.} \label{table2}
%%%%%%%%%%
\end{table}   

\subsection{Four-dimensional gaugings with $AdS_4$ and $dS_4$ vacua}
In this case, the four-dimensional gaugings can give rise to both $AdS_4$ and $dS_4$ vacua with the $dS_4$ solutions satisfying the conditions given in \eqref{eq:dS4set1}. The related five-dimensional gauge groups only admit supersymmetric $AdS_5$ vacua. All the gauge groups considered here and the associated $AdS_5$ vacua have already been studied in \cite{5D_N4_flow,5Dtwist}. 

\subsubsection{$U(1)\times SU(2)\times SU(2)$ $5D$ gauge group}
For five-dimensional gauge group $U(1)\times SU(2)\times SU(2)$, non-vanishing components of the embedding tensor can be written as 
\begin{equation}
\hat{\xi}_{12} = g_1, \qquad \hat{f}_{345} = g_2,\qquad \hat{f}_{789} = \tilde g_2\, .
\end{equation}
This gauge group arises from the four-dimensional gauge group $SO(3)^2_{+}\times SO(3)^2_-$ with the non-vanishing components of the embedding tensor given by
\begin{eqnarray}
f_{+345} &=& g_2, \qquad f_{+789} = \tilde g_2,\nonumber \\
f_{-126} &=& g_1, \qquad f_{-10,11,12} = \tilde g_1\, .
\end{eqnarray}
Therefore, we have the following identification
\begin{eqnarray}
& &\hat{f}_{\hat{M}\hat{N}\hat{P}} = f_{+\hat M\hat N\hat P}, \qquad \hat M = 3,4,5,7,8,9 \nonumber \\ 
& &\hat{\xi}_{\hat{M}\hat{N}} = f_{-\ominus \hat M \hat N}, \qquad \hat M =  1,2, \quad \ominus = 6,\oplus =12\, .
\end{eqnarray}

\subsubsection{$U(1)\times SO(3,1)$ $5D$ gauge group}
In this case, the $U(1)\times SO(3,1)$ gauge group is obtained from $SO(3)_-\times SO(3)_+\times SO(3,1)_-$ in four dimensions. The corresponding five- and four-dimensional embedding tensors are given by  
\begin{equation}
\hat{\xi}_{12} = g_1,\qquad \hat f_{345} = \hat f_{378} = -\hat f_{489} = -\hat f_{579} = g_2
\end{equation}
and
\begin{equation}
f_{-123} =g_1,\qquad f_{+456} = f_{+489}= - f_{+5,9,10} = -f_{+6,8,10} = g_2,\qquad f_{-7,11,12}=g_3
\, .
\end{equation}
Comparing these two equations, we immediately find the following identification
\begin{eqnarray}
\hat f_{\hat M\hat N\hat P} &=& f_{+\hat M+1,\hat N+1,\hat P+1}, \qquad \hat M =3,4,5,7,8,9,\nonumber \\
\hat \xi_{\hat M\hat N} &=& f_{-\ominus,\hat M+1, \hat N+1},\qquad \hat M = 1,2, \quad \ominus = 1, \oplus =12\, .
\end{eqnarray}
Note that the embedding tensor for $SO(3)_-$ has been obtained from \eqref{eq:so4em} by setting $\tilde g_1=-g_1$ together with a scaling by $\frac{1}{2\sqrt{2}}$.

\subsubsection{$U(1)\times SL(3,\mathbb R)$ $5D$ gauge group}
The $U(1)\times SL(3,\mathbb R)$ gauge group in five dimensions is gauged by the following embedding tensor
\begin{eqnarray}\label{eq:SL3o}
& &\hat \xi_{12} =g_1, \qquad \hat f_{367} = 2 g_2,\qquad  \hat f_{4,9,10}=\hat f_{5,8,10}=\sqrt{3}g_2,\nonumber \\ 
& &\qquad \hat f_{345}=\hat f_{389}=\hat f_{468}=\hat f_{497}=\hat f_{569}=\hat f_{578} = -g_2\, .
\end{eqnarray}
This gauge group can be embedded in $SO(3)_-\times SL(3,\mathbb{R})_-$ gauge group in four dimensions with the embedding tensor given by 
\begin{eqnarray}\label{eq:sl3-5d4d}
& &f_{- 123} = g_1,\qquad f_{+478} = 2 g_2,\qquad  f_{+5,10,11}= f_{+6,9,11}=\sqrt{3}g_2,\nonumber \\ 
& &\qquad f_{+456}= f_{+4,9,10}=f_{+579}=f_{+5,10,8}= f_{+6,7,10}= f_{+689} = -g_2\, .
\end{eqnarray}
We can write the relation between the embedding tensors for four- and five-dimensional gauge groups as follow 
\begin{eqnarray}\label{eq:sl3-5d4dm}
& &\hat f_{\hat M\hat N\hat P}=f_{+,\hat M+1,\hat N+1,\hat P+1},\qquad \hat M=3,4,\ldots, 10,\nonumber \\
& &\hat \xi_{\hat M\hat N}=f_{-\ominus,\hat M+1,\hat N+1},\qquad \hat M=1,2,\quad \ominus =1,\oplus =12\, .
\end{eqnarray}

\subsection{Four-dimensional gaugings with only $dS_4$ vacua}
From the results of the previous sections and in \cite{dS5}, we know that when a non-abelian compact subgroup of the R-symmetry is not gauged, $\hat{f}_{\hat m\hat n\hat p}=0$ and $f_{\alpha mnp}=0$, the gauged supergravities admit de Sitter vacua. Both in four and five dimensions, these gauge groups take the form of a product of non-compact groups. We will give some relations between this type of gaugings in four and five dimensions. 

\subsubsection{$SO(1,1)\times SU(2,1)$ $5D$ gauge group}
The five-dimensional gauge group $SO(1,1)\times SU(2,1)$ corresponds to the following embedding tensor
\begin{eqnarray}
\hat \xi_{5,10} &=&g_1, \nonumber \\
\hat f_{129}&=&\hat f_{138} = \hat f_{147} = \hat f_{248} = -\hat f_{349} = -\hat f_{237} =g_2,\nonumber \\ 
\hat f_{789} &=& -2g_2, \qquad \hat f_{346} = \hat f_{126} = \sqrt{3}g_2\, .
\end{eqnarray}
This is identified with $SO(2,1)_+\times SU(2,1)_+$ gauge group in four dimensions with the embedding tensor
\begin{eqnarray}
f_{+\, 129}  &=& f_{+\,138}=f_{+\,147} =  f_{+\,248} =-f_{+\,237}=-f_{+\,349} =g_2,\nonumber \\ 
f_{+\,789} &=& -2g_2, \qquad f_{+\,126} = f_{+\,346}=\sqrt{3}g_2,\qquad f_{-\,5,6,11} = g_1\, .\label{SU2_1_1}
\end{eqnarray}
In this case, it is straightforward to see the relation between the two gauge groups 
\begin{eqnarray}
& &\hat f_{\hat M\hat N\hat P} = f_{+\hat M \hat N\hat P}, \qquad \hat M= 1,2,3,4,7,8,9,\nonumber \\
& &\hat \xi_{\hat M\hat N} = f_{-\ominus,\hat M+1,\hat N+1}, \qquad \hat M =  5,10,  \quad \ominus = 5, \oplus =12\, .
\end{eqnarray}

\subsubsection{$SO(1,1)\times SO(2,1)$ $5D$ gauge group}		
As shown in \cite{dS5}, $SO(1,1)\times SO(2,1)$ gauge group can be embedded in $SO(5,n)$ in different forms. These forms depend on how the $SO(1,1)$ factor is gauged. In general, the $SO(1,1)$ can be embedded as a diagonal subgroup of $d$ $SO(1,1)$ factors, $SO(1,1)^{(d)}_{\textrm{diag}}\sim [SO(1,1)^1\times \ldots\times SO(1,1)^d]_{\textrm{diag}}$ in the notation of \cite{dS5}. It has also been shown in \cite{dS5} that, due to the presence of a nonabelian non-compact factor, there are only three possibilities with $d=1,2,3$. For a simple embedding as $SO(1,1)\times SO(2,1)$ gauge group, with $SO(1,1)^{(1)}$ simply denoted by $SO(1,1)$, the embedding tensor is given by
\begin{equation}
\hat \xi_{56} = g_1, \qquad \hat f_{237} = g_2\, .
\end{equation}
We identify this gauge group as arising from the $SO(2,1)_+\times SO(2,1)_+$ in four dimensions with the embedding tensor
\begin{equation}
f_{+237} = g_2, \qquad f_{-5,6,10} = g_1\, .
\end{equation}
This embedding tensor has been obtained from \eqref{eq:so22em} by setting $\tilde g_1=g_1$ and $\tilde g_2=g_2$ and renaming $\sqrt{2}g_{1,2}\rightarrow g_{2,1}$. We then see the following relation between the two embedding tensors 
\begin{eqnarray}
& &\hat f_{\hat M\hat N\hat P} = f_{+\hat M\hat N\hat P}, \qquad \hat M=2,3,7,\nonumber \\
& &\hat \xi_{\hat M\hat N} = f_{-\ominus \hat M\hat N},\qquad \hat M =  5,6, \quad \ominus = 10,\oplus =4\, .
\end{eqnarray}
It should be pointed out that, in both five and four dimensions, there are additional $SO(2)$ factors under which matter fields are not charged. These abelian factors correspond to the gauge fields not participating in the above gauge groups. Furthermore, the four-dimensional gauge group $SO(2,1)_+\times SO(2,1)_+$ has not been separately listed in \cite{de-Roo-Panda-2}. However, this gauge group can be obtained as a particular subgroup of either $SO(2,2)_+\times SO(2,2)_-$ or $SO(2,1)_+^3\times SO(3)_+$ gauge groups.  

\subsubsection{$SO(1,1)^{(2)}_{\textrm{diag}}\times SO(2,1)$ $5D$ gauge group}		
We then move to $SO(1,1)^{(2)}_{\textrm{diag}}\times SO(2,1)$ gauge group with the embedding tensor
\begin{equation}		
\hat \xi_{18}=\hat \xi_{27}=g_1,\qquad \hat f_{4,5,10}=g_2\, .		
\end{equation}		
This gauge group is related to $SO(3,1)_+\times SO(2,1)_+\times SO(2,1)_-$ gauge group in four dimensions with the embedding tensor
\begin{equation}
f_{-\, 789}  = -f_{-\,129}=-f_{-\,138} = f_{-\,723} =-g_1,\qquad f_{+\,5,6,11} = g_2,\qquad f_{+\, 4,10,12}=g_3
\end{equation}
by the following relation
\begin{eqnarray}
& &\hat f_{\hat M\hat N\hat P} = f_{+\hat M+1,\hat N+1,\hat P+1}, \qquad \hat M=4,5,10,\nonumber \\
& &\hat \xi_{\hat M\hat N} = f_{-\ominus, \hat M+1,\hat N+1},\qquad \hat M =  1,2,7,8, \quad \ominus = 1,\oplus =12\, .
\end{eqnarray}			

\subsubsection{$SO(1,1)^{(3)}_{\textrm{diag}}\times SO(2,1)$ $5D$ gauge group}			
We now consider the final form of $SO(1,1)\times SO(2,1)$ gauge group in five dimensions namely $SO(1,1)^{(3)}_{\textrm{diag}}\times SO(2,1)$ with the following embedding tensor
\begin{equation}		
\hat \xi_{18}=\hat \xi_{27}=\hat \xi_{36}=g_1,\qquad \hat f_{4,5,10}=g_2\, .		
\end{equation}		
This gauge group is related to $SO(2,1)_+\times SU(2,1)_+$ gauge group in four dimensions with the embedding tensor
\begin{eqnarray}
f_{-\, 129}  &=& f_{-\,138}=f_{-\,147} =  f_{-\,248} =-f_{-\,237}=-f_{-\,349} =g_1,\nonumber \\ 
f_{-\,789} &=& -2g_1, \qquad f_{-\,1,2,10} = f_{-\,3,4,10}=\sqrt{3}g_1,\qquad f_{+\,5,6,11} = g_2\label{SU2_1_2}
\end{eqnarray}
by the following relation
\begin{eqnarray}
& &\hat f_{\hat M\hat N\hat P} = f_{+\hat M+1,\hat N+1,\hat P+1}, \qquad \hat M=4,5,10,\nonumber \\
& &\hat \xi_{\hat M\hat N} = f_{-\ominus, \hat M+1,\hat N+1},\qquad \hat M =  1,2,3,6,7,8, \quad \ominus = 1,\oplus =12\, .
\end{eqnarray}
It should be noted that the embedding tensor \eqref{SU2_1_2} is just the one in \eqref{SU2_1_1} with $+\rightarrow -$ and $g_2\rightarrow -g_1$\, .

\subsubsection{$SO(1,1)\times SO(2,2)$ $5D$ gauge group}		
In this case, the five-dimensional gauge group $SO(1,1)\times SO(2,2)\sim SO(1,1)\times SO(2,1)\times SO(2,1)$ is gauged by the embedding tensor
\begin{equation}
\hat \xi_{5,10} = g_1, \qquad \hat f_{237} = g_2,\qquad \hat f_{+189} = g_3\, .
\end{equation}
We identify this gauge group as related to $SO(2,1)_+^2\times SO(2,1)^2_-$ gauge group in four dimensions with the embedding tensor 	
\begin{equation}
f_{+237} = g_2,\qquad f_{+189} = g_3, \qquad f_{-5,6,10} = -g_1,\qquad f_{-4,11,12} = g_4
\end{equation}
by the following relation
\begin{eqnarray}
& &\hat f_{\hat M \hat N \hat P} = f_{+\hat M \hat N \hat P},\qquad  \hat M = 1,8,9,2,3,7,\nonumber \\
& &\hat \xi_{\hat M \hat N} = f_{-\ominus \hat M \hat N},\qquad \hat M = 5,10,\quad \ominus =6,\oplus =12\, .
\end{eqnarray}
It should be noted that the second $SO(2,1)$ factor in this case has the compact $SO(2)$ subgroup along the R-symmetry direction. This $SO(2,1)$ is called $SO(2,1)'$ in \cite{dS5} to distinguish it from the other $SO(2,1)$ factor with the compact part along the matter direction.
		
\subsubsection{$SO(1,1)\times SO(3,1)$ $5D$ gauge group}
In this case, the five-dimensional gauge group is gauged by the following embedding tensor
\begin{equation}
\hat \xi_{5,10}  =g_1,\qquad \hat f_{789} =-\hat f_{129} = -\hat f_{138} = \hat f_{237}= g_2\, .
\end{equation}
This gauge group is obtained from $SO(3,1)_+\times SO(2,1)_+\times SO(2,1)_-$ gauge group in four dimensions with the embedding tensor
\begin{equation}
f_{-4,6,11}= -g_1,\qquad f_{+789} =-f_{+129} = -f_{+138} = f_{+237} =  g_2,\qquad  f_{-5,7,12}=g_3\, .
\end{equation}
The relation between the two embedding tensors is then given by
\begin{eqnarray}
& &\hat f_{\hat M\hat N\hat P} = f_{+\hat M\hat N\hat P}, \qquad \hat M =1,2,3,7,8,9, \nonumber \\ 
& &\hat \xi_{\hat M\hat N} = f_{-\ominus,\hat M+1,\hat N+1},\qquad \hat M= 5,10, \quad \ominus = 6,\oplus =12\, .
\end{eqnarray}

\subsubsection{$SO(1,1)^{(2)}_{\textrm{diag}}\times SO(3,1)$ $5D$ gauge group}
As shown in \cite{dS5}, there is another embedding of the five-dimensional $SO(1,1)\times SO(3,1)$ gauge group in the form of $SO(1,1)^{(2)}_{\textrm{diag}}\times SO(3,1)$ with the embedding tensor
\begin{equation}
\hat \xi_{29} =\hat \xi_{38} =g_1,\qquad \hat f_{789} =-\hat f_{129} = -\hat f_{138} = \hat f_{237}= g_2\, .
\end{equation}
We identify that this gauge group is related to $SO(3,1)_+\times SO(3,1)_+$ gauge group in four dimensions with the embedding tensor
\begin{eqnarray}
& &f_{-789} = -f_{-129} = -f_{-138} = f_{-723} = -g_1,\nonumber \\ 
& &f_{+10,11,12} = -f_{+4,5,12} = -f_{+4,6,11} =f_{+10,5,6} = g_2
\end{eqnarray}
via the following relation
\begin{eqnarray}
& &\hat f_{\hat M\hat N\hat P} = f_{+\hat M+3,\hat N+3\hat P+3}, \qquad \hat M =1,2,3,7,8,9, \nonumber \\ 
& &\hat \xi_{\hat M\hat N} = f_{-\ominus\hat M\hat N},\qquad \hat M= 2,3,8,9, \quad \ominus = 1,\oplus =7\, .
\end{eqnarray}

\subsubsection{$SO(1,1)\times SO(4,1)$ $5D$ gauge group}
To gauge this group, we need to couple the five-dimensional $N=4$ supergravity to at least $\hat n=7$ vector multiplets. The embedding tensor is given by 
\begin{eqnarray}
& &\hat \xi_{5,12} = g_1,\qquad \hat f_{126} = \hat f_{137} =\hat f_{149} =\hat f_{238} = \hat f_{2,4,10} =\hat f_{3,4,11} =  g_2,\nonumber \\
& & \hat f_{678} = \hat f_{6,9,10} = \hat f_{7,9,11} = \hat f_{8,10,11} = -g_2\, .\qquad\,\,\,
\end{eqnarray}
There is no known four-dimensional gauging that can be identified with this gauge group upon a circle reduction. We will use the relations given in \eqref{eq:4d5d} to construct the following four-dimensional embedding tensor
\begin{eqnarray}
& & f_{+MNP}=\hat f_{M-1,N-1,P-1}, \qquad M =2,3,4,5,7,\ldots ,12, \nonumber \\ 
& &f_{-\ominus M N}=\hat \xi_{ M-1,N-1},\qquad M= 6,13, \quad \ominus = 1\, .
\end{eqnarray}
The result is given by
\begin{eqnarray}\label{eq:4dso41}
& &f_{+237} = f_{+248} =f_{+2,5,10} = f_{+349} = f_{+3,5,11} =f_{+4,5,12} = g_2, \nonumber \\
& &
f_{+789} = f_{+7,10,11} = f_{+8,10,12} = f_{+9,11,12} = -g_2,\qquad 
f_{-1,6,13} = g_1\, .
\end{eqnarray}
This corresponds to $SO(2,1)_+\times SO(4,1)_+$ gauge group. In this case, the four-dimensional gauged supergravity also needs to couple to at least $7$ vector multiplets.  
\\
\indent It is now straightforward to compute the scalar potential and determine the critical points. With only the dilaton and axion non-vanishing, the result is given by
\begin{eqnarray}
V=\frac{1}{2} e^{-\phi } \left[e^{2 \phi } \left(g_1^2 \chi ^2+6 g_2^2\right)+g_1^2\right]
\end{eqnarray}
with a critical point at
\begin{eqnarray}
\chi = 0, \qquad \phi = \ln\left[\pm \frac{g_1}{\sqrt{6} g_2}\right].
\end{eqnarray}
After setting $g_2 = \pm \frac{1}{\sqrt{6}}g_1$, we obtain a $dS_4$ vacuum with $V_0 = g_1^2$ and satisfying \eqref{eq:dS4set2}. We also note that the electric-magnetic dual of \eqref{eq:4dso41}, with $-$ and $+$ components interchanged, leads to the same potential and critical point. 
\\
\indent Finally, scalar masses at this critical point are given by
\begin{equation}
m^2L^2=6_{\times 2},\quad \frac{3}{2}(1+2\sqrt{2})_{\times 4},\quad \frac{3}{2}(1-2\sqrt{2})_{\times 4},\quad 0_{\times 6},\quad 3_{\times 12},\quad \left[\frac{3}{2}\right]_{\times 16} 
\end{equation}
where we have used the $dS_4$ radius $L=\frac{3}{V_0}$. This $dS_4$ vacuum is unstable due to the negative mass value $\frac{3}{2}(1-2\sqrt{2})$. The mass value $m^2L^2=6$ corresponds to that of the dilaton and axion. The six massless scalars correspond to the Goldstone bosons of the symmetry breaking $SO(2,1)\times SO(4,1)\rightarrow SO(2)\times SO(4)$.

%%%%%%%%%%%%%%%%%%%%%%%%%%%%%%%%%%%%%%%%%%%%%%%%%%%%%%%%%%%
\section{Conclusions}\label{conclusions}
In this paper, we have studied $dS_4$ vacua of four-dimensional $N=4$ gauged supergravity coupled to vector multiplets. By requiring that the scalar potential is extremized and positive, we have derived a set of conditions for determining a general form of gauge groups admitting $dS_4$ vacua by adopting a simple ansatz. This extends the previous result in five-dimensional $N=4$ gauged supergravity and provides a useful approach for finding $dS_4$ vacua in $N=4$ gauged supergravity. We have also given some relations between the embedding tensors of four- and five-dimensional gauge groups that could be related by a simple circle reduction. From this analysis, we have given a new example of four-dimensional gauge group, $SO(2,1)\times SO(4,1)$, that gives a $dS_4$ vacuum. This has not previously been studied since the gauging requires the coupling to at least seven vector multiplets.  
\\
\indent Unlike in five dimensions, we find two large classes of gauge groups that give $dS_4$ vacua as maximally symmetric backgrounds of the matter-coupled $N=4$ gauged supergravity. For the first class, the gauge groups take a general form of $G_{\textrm{e}}\times G_{\textrm{m}}\times G_0^v$ in which $G_{\textrm{e(m)}}$ is electrically (magnetically) gauged and contains an $SO(3)$ subgroup. $G_0^v$ is a compact group gauged by vector fields in the vector multiplets. These gauge groups are precisely the ones that lead to supersymmetric $AdS_4$ vacua studied in \cite{AdS4_N4_Jan}. Two different types of vacua, $AdS_4$ and $dS_4$, arise from different coupling ratios between $G_{\textrm{e}}$ and $G_{\textrm{m}}$ factors. This result is obtained from imposing the conditions that $\langle A^{ij}_1\rangle =\langle {A_{2ai}}^j\rangle=0$. These conditions take a very similar form to those for the existence of supersymmetric $AdS_4$ vacua. We have explicitly verified that the potential is extremized by these conditions.  
\\
\indent For the second class, we have imposed another set of conditions, $\langle A^{ij}_1\rangle=\langle A^{ij}_2\rangle=0$, and found that the gauge groups generally take the form $SO(2,1)\times SO(2,1)' \times G_{\textrm{nc}}\times G'_{\textrm{nc}}\times H_{\textrm{c}}$. $G_{\textrm{nc}}$ and $G'_{\textrm{nc}}$ are non-compact groups with the compact parts embedded in the matter directions while the compact $SO(2)\times SO(2)'\subset SO(2,1)\times SO(2,1)'$ is embedded along the R-symmetry directions. As in the previous case, $SO(2,1)\times G_{\textrm{nc}}$ ($SO(2,1)'\times G'_{\textrm{nc}}$) is electrically (magnetically) gauged, and $H_{\textrm{c}}$ is a compact group. It should be emphasized that only $G_{\textrm{nc}}$ and $G'_{\textrm{nc}}$ are necessary for the $dS_4$ vacua to exist. 
\\
\indent Given that our ansatz is rather simple, it is remarkable that the above results encode all semi-simple gauge groups that are previously known to give $dS_4$ vacua of $N=4$ gauged supergravity. Two different sets of these gauge groups have also been noted in \cite{de-Roo-Panda-2}, and these correspond to the two sets of conditions given in this paper. The results given here are hopefully useful for finding $dS_4$ vacua and could be interesting in the dS/CFT correspondence and cosmology. 
\\
\indent In this paper, we have looked at only semisimple gauge groups. It is also interesting to consider non-semisimple gauge groups listed in \cite{CSO_N4_4D} and those arising from flux compactification studied in \cite{Dibitetto_dS} and \cite{Dibitetto_dS2}. In deriving all the conditions for the existence of $dS_4$ vacua, we have not restricted the gauge groups to be semisimple. Therefore, our conditions are also valid for non-semisimple gauge groups. In particular, it can be verified that, for the $dS_4$ vacuum from $ISO(3)\times ISO(3)$ gauge group considered in \cite{Dibitetto_dS2}, we have $\langle A^{ij}_1\rangle=\langle {A_{2ai}}^j\rangle =0$. This $dS_4$ solution is accordingly of the first type described by the criteria given in \eqref{eq:dS4set1}. A systematic classification of non-semisimple groups leading to $dS_4$ vacua in $N=4$ gauged supergravity is worth considering. 
\\
\indent Given the success in $N=4$ gauged supergravities in both four and five dimensions, it is natural to extend this approach to other gauged supergravities with different numbers of supersymmetries in various dimensions. The success of this approach also suggests that the conditions we have derived might have deeper meaning although they are originally obtained from a simple assumption. It would be of particular interest to have a definite conclusion whether there is some explanation for these conditions within gauged supergravity and string/M-theory or these conditions are just a tool for finding de Sitter solutions.    
\vspace{0.5cm}\\
%%%%%%%%%%%%%%%%%%%%%%%%%%%%%%%%%%%%%%%%%%%%%%%%%%%%%%%%%%%%%%%%%%
{\large{\textbf{Acknowledgement}}} \\
P. K. is supported by The Thailand Research Fund (TRF) under grant RSA6280022.
%%%%%%%%%%%%%%%%%%%%%%%%%%%%%%%%%%%%%%%%%%%%%%%%%%%%%%%%%%%%%%%%%%  

\appendix
\section{Useful formulae}
In this appendix, we collect some useful identities involving $SO(6)$ gamma matrices which are useful in the analysis of the constraints on the embedding tensor. This appendix closely follows the dicussion in \cite{AdS4_N4_Jan} and \cite{meta-stable}. The $8\times 8$ gamma matrices of $SO(6)$, ${\gamma_{mI}}^I$, $I,J=1,2,\ldots, 8$, satisfy the Clifford algebra
\begin{equation}
{\gamma_{mI}}^K{\gamma_{nK}}^J+{\gamma_{nI}}^K{\gamma_{mK}}^J=2\delta^J_I\delta_{mn}\, .
\end{equation}
${\gamma_{mI}}^J$ can be written in terms of the chirally projected $4\times 4$ gamma matrices $\Gamma^{ij}_m$, $i,j=1,2,3,4$, and their complex conjugate ${\Gamma_{mij}}=(\Gamma_m^{ij})^*=\frac{1}{2}\epsilon_{ijkl}\Gamma^{kl}_m$ as follow
\begin{equation}
{\gamma_{mI}}^J=\begin{pmatrix}
0 & \Gamma_m^{ij}\\
\Gamma_{mij}& 0 
\end{pmatrix} \, .
\end{equation}
$\Gamma_m$ satisfy the Clifford algebra
\begin{equation}
\{\Gamma_m,\Gamma_n^*\}=2\delta_{mn}\mathbf{I}_4\, .
\end{equation}
An explicit form of these matrices can be chosen as
\begin{eqnarray}
\Gamma_1&=&i\mathbf{I}_2\otimes \sigma_2,\qquad \Gamma_2=i\sigma_2\otimes \sigma_3,\qquad \Gamma_3=i\sigma_2\otimes \sigma_1,\nonumber \\
\Gamma_4&=&-\sigma_3\otimes \sigma_2,\qquad \Gamma_5=-\sigma_2\otimes \mathbf{I}_2,\qquad \Gamma_6=-\sigma_1\otimes \sigma_2\, .
\end{eqnarray}
\indent The $SO(6)$ generators in the chiral spinor representation or $SU(4)$ generators in the fundamental representation are given by
\begin{equation}
{(\Gamma_{mn})^i}_j=\frac{1}{2}\Gamma^{ik}_m(\Gamma^*_n)_{kj}\, .
\end{equation}
The other antisymmetric products satisfy
\begin{eqnarray}
(\Gamma_{mnp})^{ij}=\Gamma^{ik}_{[m}\Gamma_{nkl}\Gamma_{p]}^{lj}&=&i\epsilon_{mnpqrs}\Gamma^{ik}_q\Gamma_{rkl}
\Gamma^{lj}_s,\label{3_Gamma}\\
{(\Gamma_{mnpq})^i}_j=\Gamma^{ik_1}_{[m}\Gamma_{nk_1k_2}\Gamma_{p}^{k_2k_3}\Gamma_{q]k_3j}&=&i\epsilon_{mnpqrs}
\Gamma^{ik}_r\Gamma_{skj},\\
\Gamma^{ik_1}_{[m}\Gamma_{nk_1k_2}\Gamma^{k_2k_3}_{p}\Gamma_{qk_3k_4}\Gamma^{k_4k_5}_{r}\Gamma_{s]k_5j}
&=&i\delta^i_j\epsilon_{mnpqrs}\, .
\end{eqnarray}
Some useful identities are given by 
\begin{eqnarray}
\{\Gamma_{mn},\Gamma_{pq}\}&=&2\Gamma_{mnpq}+2\delta_{np}\delta_{mq}-2\delta_{mp}\delta_{nq},\label{Gamma_mn_anti_com}\\
\textrm{Tr}(\Gamma_{mnp}\Gamma_{qrs})&=&-4\delta_{mq}\delta_{nr}\delta_{ps}+4\delta_{mq}\delta_{ns}\delta_{pr}
+4\delta_{mr}\delta_{nq}\delta_{ps}\nonumber \\
& &-4\delta_{mr}\delta_{ns}\delta_{pq}-4\delta_{ms}\delta_{nq}\delta_{pr}+4\delta_{ms}\delta_{nr}\delta_{pq}\, .\label{6_Gamma_trace}
\end{eqnarray}

%%%%%%%%%%%%%%%%%%%%%%%%%%%%%%%%%%%%%%%%%%%%%%%%%%%%%%%%%%%%%%%%%%%%%%%%%%%%%%%%%%%%%%%%%%%%%%%%%%%%%%%%%%%%%%%%%%%%%%%%%%%%%%%%%%%%%%%%%%%%%%%%%%%%%%%%%%%%%%%%


\begin{thebibliography}{99}
\bibitem{dS_cosmo1} C. L. Bennet et al., ``First Year Wilkinson Microwave Anisotropy Probe (WMAP)
Observations: Preliminary Maps and Basic Results'', Astrophysics. J. Suppl. 148 (2003) \textbf{1}, arXiv: astro-ph/0302207.
\bibitem{dS_cosmo2} A. G. Riess et al., ``Observational Evidence from Supernovae for an Accelerating Universe and
a Cosmological Constant'', Astron. J. 116 (1998) \textbf{1009}, arXiv: astro-ph/9805201.
\bibitem{dS_cosmo3} S. Perlmutter et al., ``Measurement of Omega and Lambda from 42 High-Redshift Supernovae'',
Astrophys. J. 517 (1999) \textbf{565}, arXiv: astro-ph/9812133.
\bibitem{dS_CFT} A. Strominger, ``The dS/CFT Correspondence'', JHEP 10 (2001) \textbf{034}, arXiv: hep-th/0106113.
\bibitem{maldacena} J. M. Maldacena, ``The large $N$ limit of
superconformal field theories and supergravity'', Adv. Theor. Math.
Phys. \textbf{2} (1998) 231-252, arXiv: hep-th/9711200.
\bibitem{RS} L. Randall and R. Sundrum, ``An Alternative to compactification,” Phys. Rev. Lett. 83 (1999) 4690–4693, arXiv: hep-th/9906064.
\bibitem{dS1} G. W. Gibbons and C. M. Hull, ``De Sitter space from warped supergravity solutions'', hep-th/0111072.
\bibitem{dS2} I. P. Neupane, ``De Sitter brane-world, localization of gravity, and the cosmological constant'', Phys. Rev. \textbf{D83} (2011) 086004, 1011.6357.
\bibitem{dS3} M. Minamitsuji and K. Uzawa, ``Warped de Sitter compactifications'', JHEP 01 (2012) 142, arXiv: 1103.5326.
\bibitem{dS4} M. Dodelson, X. Dong, E. Silverstein, and G. Torroba, ``New solutions with accelerated expansion in string theory'', JHEP 12 (2014) \textbf{050}, arXiv: 1310.5297.
\bibitem{flux_moduli} B. de Carlos, A. Guarino, J. M. Moreno, ``Flux moduli stabilisation, Supergravity algebras and no-go theorems'', JHEP 01 (2010) \textbf{012}, arXiv: 0907.5580.
\bibitem{complete_Mink_vacua} B. de Carlos, A. Guarino, J. M. Moreno, ``Complete classification of Minkowski vacua in generalised flux models'', JHEP 02 (2010) \textbf{076}, arXiv: 0911.2876.
\bibitem{dS_string6} J. Blaback, U. Danielsson, and G. Dibitetto, ``Fully stable dS vacua from generalised fluxes'', JHEP 08 (2013) \textbf{054}, arXiv:1301.7073.
\bibitem{dS6} C. Damian, L. R. Diaz-Barron, O. Loaiza-Brito, and M. Sabido, ``Slow-Roll Inflation in Non-geometric Flux Compactification'', JHEP 06 (2013) \textbf{109}, arXiv: 1302.0529.
\bibitem{dS_Sdual} C. Damian and O. Loaiza-Brito, ``More stable de Sitter vacua from S-dual nongeometric fluxes'', Phys. Rev. \textbf{D88} (2013) 046008, arXiv: 1304.0792.
\bibitem{dS8} R. Blumenhagen, C. Damian, A. Font, D. Herschmann, and R. Sun, ``The Flux-Scaling Scenario: De Sitter Uplift and Axion Inflation'', Fortsch. Phys. \textbf{64} (2016), 536–550, arXiv: 1510.01522.
\bibitem{dS9} K. Dasgupta, G. Rajesh, and S. Sethi, ``M theory, orientifolds and G - flux,” JHEP \textbf{08} (1999) 023, arXiv: hep-th/9908088.
\bibitem{dS_string1} G. W. Gibbons, Aspects of supergravity theories, in Supersymmetry, Supergravity and Related Topics, editors F. Del Aguila, J. A. de Azc´arraga and L. E. Ib´a˜nez, World Scientific 1985.
\bibitem{dS_string2} B. de Wit, D.J. Smit and N.D. Hari Dass, Residual supersymmetry of compactified D = 10
supergravity, Nucl. Phys. \textbf{B283} (1987) 165.
\bibitem{dS_string3} J. Maldacena and C. Nunez, Supergravity description of field theories on curved manifolds and a no go theorem, Int. J. Mod. Phys. \textbf{A16} (2001) 822, hep-th/0007018.
\bibitem{dS_string3_1} M. P. Hertzberg, S. Kachru, W. Taylor, and M. Tegmark, ``Inflationary Constraints on Type IIA String Theory'', JHEP 0712 (2007) \textbf{095}, arXiv:0711.2512.
\bibitem{dS_string4} P. K. Townsend, ``Cosmic acceleration and M-theory'', hep-th/0308149.
\bibitem{dS_string5} U. Danielsson and G. Dibitetto, ``On the distribution of stable de Sitter vacua'', JHEP 03 (2013) \textbf{018}, arXiv:1212.4984.
\bibitem{Danielsson-18} U. Danielsson and T. van Riet, ``What if string theory has no de Sitter vacua?'', Int. J. Mod. Phys. \textbf{D27} (2018) 12, 1830007, arXiv: 1804.01120. 
\bibitem{Hull-84} C. M. Hull, ``Noncompact gaugings of $N=8$ supergravity'', Phys. Lett. \textbf{B142} (1984) 39.
\bibitem{de-Roo-Panda-1} M. de Roo, D. B. Westra and S. Panda, ``De Sitter solutions in $N = 4$ matter coupled supergravity'', JHEP 02 (2003) \textbf{003}, arXiv: hep-th/0212216.
\bibitem{Trigiante} P. Fre, M. Trigiante and A. Van Proeyen, "Stable de Sitter vacua from $N = 2$ supergravity'', Class. Quant. Grav. \textbf{19} (2002) 4167–4194, arXiv:hep-th/0205119.
\bibitem{de-Roo-Panda-2} M. de Roo, D. B. Westra, S. Panda and M. Trigiante, ``Potential and mass-matrix in gauged N = 4 supergravity'', JHEP 11 (2003) \textbf{022}, arXiv:hep-th/0310187.
\bibitem{Roest-09} D. Roest and J. Rosseel, ``de-Sitter in extended gauged supergravity'', Phys. Lett. \textbf{B685} (2010) 201-207, arXiv: hep-th/0912.4440.
\bibitem{smet-PhD} G. Smet, PhD thesis, ``de Sitter space and supergravity in various dimensions" (2006).
\bibitem{gunaydin-00} M. Gunaydin and M. Zagermann, ``The vacua of 5d, $N = 2$ gauged Yang-Mills/Einstein/tensor supergravity: Abelian case'', Phys. Rev. \textbf{D62} (2000) 044028 arXiv: hep-th/0002228.
\bibitem{ogetbil-1} O. Ogetbil, ``A General Study of Ground States in Gauged $N = 2$ Supergravity Theories with Symmetric Scalar Manifolds in 5 Dimensions", arXiv: hep-th/0612145.
\bibitem{ogetbil-2} O. Ogetbil, ``Stable de Sitter Vacua in 4 Dimensional Supergravity Originating from 5 Dimensions", Phys. Rev. \textbf{D78} 105001, arXiv: 0809.0544.
\bibitem{smet-05}  B. Cosemans and G. Smet, ``Stable de Sitter vacua in $N = 2$, $D = 5$ supergravity", Class. Quant. Grav. \textbf{22} (2005) 2359–2380 arXiv: hep-th/0502202.
\bibitem{dS7} G. Dibitetto, J. J. Fernandez-Melgarejo and D. Marques, ``All gaugings and stable de Sitter in $D = 7$ half-maximal supergravity'', JHEP 11 (2015) \textbf{037}, arXiv: 1506.01294.
\bibitem{dS5} H. L. Dao and P. Karndumri, ``$dS_5$ vacua from matter-coupled $5D$ $N=4$ gauged supergravity'', arXiv:1906.09776.
\bibitem{N4_gauged_SUGRA} J. Schon and M. Weidner, ``Gauged $N=4$ supergravities'', JHEP 05 (2006) \textbf{034}, arXiv: hep-th/0602024.
\bibitem{Eric_N4_4D} E. Bergshoeff, I. G. Koh and E. Sezgin, ``Coupling of Yang-Mills to $N=4$, $d=4$ supergravity'', Phys. Lett. \textbf{B155} (1985) 71-75.
\bibitem{de_Roo_N4_4D} M. de Roo and P. Wagemans, ``Gauged matter coupling in $N=4$ supergravity'', Nucl. Phys. \textbf{B262} (1985) 644-660.
\bibitem{AdS4_N4_Jan} J. Louis and H. Triendl, ``Maximally supersymmetric $AdS_4$ vacua in $N=4$ supergravity'', JHEP 10 (2014) \textbf{007}, arXiv:1406.3363.
\bibitem{4D_N4_flows} P. Karndumri and K. Upathambhakul, "Holographic RG flows in N = 4 SCFTs from half-maximal gauged supergravity", Eur. Phys. J. \textbf{C78} (2018) 626, arXiv:hep-th/1806.01819.
\bibitem{5D_N4_flow} H. L. Dao and P. Karndumri, ``Holographic RG flows and $AdS_5$ black strings from 5D half-maximal gauged supergravity'', Eur. Phys. J. \textbf{C79} (2019) 137, arXiv: 1811.01608.
\bibitem{5Dtwist} H. L. Dao and P. Karndumri, ``Supersymmetric $AdS_5$ black holes and strings from 5D $N=4$ gauged supergravity'', Eur. Phys. J. \textbf{C79} (2019) 247, arXiv: 1812.10122.
\bibitem{CSO_N4_4D} M. de Roo, D. B. Westra and S. Panda, ``Gauging CSO groups in $N=4$ Supergravity'', JHEP 09 (2006) \textbf{011}, arXiv: hep-th/0606282.
\bibitem{Dibitetto_dS} G. Dibitetto, R. Linares and D. Roest, ``Flux Compactifications, Gauge Algebras and De Sitter'', Phys. Lett. \textbf{B688} (2010) 96-100, arXiv: 1001.3982.
\bibitem{Dibitetto_dS2} G. Dibitetto, A. Guarino and D. Roest, ``Charting the landscape of $N=4$ flux compactifications'', JHEP 03 (2011) \textbf{137}, arXiv: 1102.0239.
\bibitem{meta-stable} A. Borghese and D. Roest, ``Metastable supersymmetry breaking in extended supergravity'', JHEP 05 (2011) \textbf{102}, arXiv: 1012.3736.
\end{thebibliography}
\end{document}